\documentclass[12pt]{article}
\pdfoutput=1
\usepackage[a4paper]{geometry}
\usepackage{jheppub, amsmath,amssymb,amsfonts,amsxtra, mathrsfs, makeidx,graphics,graphicx,amsthm,epsfig, ytableau,bm,longtable,float, color,tikz,mathtools,xfrac,footnote,rotating, lscape, makecell, environ,mathtools, empheq, colortbl}

\usetikzlibrary{positioning}
\usetikzlibrary{chains}
\usetikzlibrary{arrows,fit,decorations.pathreplacing}
\tikzstyle{every picture}+=[remember picture]
\tikzstyle{na} = [baseline]

\usetikzlibrary{arrows, decorations.markings, calc, fadings, decorations.pathreplacing, patterns, decorations.pathmorphing, positioning}

\def\node#1#2{\overset{#1}{\underset{#2}{{\color{gray} \bullet}}}}

\def\node#1#2{\overset{#1}{\underset{#2}{\circ}}}

\tikzstyle{every picture}+=[remember picture]
\tikzstyle{na} = [baseline=-.5ex]

\usepackage{bbm}
\usepackage{subfigure}

\numberwithin{equation}{section}

\newcommand{\be}{\begin{equation}} \newcommand{\ee}{\end{equation}}
\newcommand{\bea}{\begin{equation} \begin{aligned}} \newcommand{\eea}{\end{aligned} \end{equation}}

\def\tilde{\widetilde}

\def\rt2{\sqrt{2}}

\def\Tr{\mathop{\rm Tr}}

\def\CT{{\cal T}}

\def\CW{{\cal W}}

\def\1{{\ds 1}}

\newcommand{\cB}{\mathcal{B}}
\newcommand{\cC}{\mathcal{C}}

\newcommand{\cM}{\mathcal{M}}
\newcommand{\cN}{\mathcal{N}}

\newcommand{\cT}{\mathcal{T}}

\newcommand{\cV}{\mathcal{V}}
\newcommand{\cW}{\mathcal{W}}

\def\repa{\raise4pt\hbox{$\square$}\mkern-14mu\raise-4pt\hbox{$\square$}}
\def\repab{\overline{\raise4pt\hbox{$\square$}\mkern-14mu\raise-4pt\hbox{$\square$}\mkern-1mu}}

\def\smileface{\ensuremath{\hbox{\large$\bigcirc$}\mkern-15mu\raise-1pt\hbox{\scriptsize$\smallsmile$}%
\mkern-10mu\raise4pt\hbox{..}\mkern4mu}}
\def\frownface{\ensuremath{\hbox{\large$\bigcirc$}\mkern-15mu\raise-1pt\hbox{\scriptsize$\smallfrown$}%
\mkern-10mu\raise4pt\hbox{..}\mkern4mu}}

\newcommand{\ba}{\begin{array}}
\newcommand{\ea}{\end{array}}
\newcommand{\bi}{\begin{itemize}}
\newcommand{\ei}{\end{itemize}}
\def\vec#1{\bm{#1}}
\def\bea#1\eea{\allowdisplaybreaks \begin{align}#1\end{align}}
 \newcommand{\ben}{\begin{enumerate}}
\newcommand{\een}{\end{enumerate}}
\newcommand{\bean}{\begin{eqnarray*}}
\newcommand{\eean}{\end{eqnarray*}}

\newcommand{\comment}[1]{}

\newcommand{\bS}{\mathbb{S}}

\definecolor{light-gray}{gray}{0.7}

\def\aup#1 {\overset{#1}{\uparrow} \, \overset{\tilde{#1}}{\downarrow}}

\tikzset{snake it/.style={decorate, decoration={snake, amplitude=.4mm, segment length=2mm,
                       post length=0mm,pre length=0mm}}}

\newcommand{\bpic}{\begin{tikzpicture}}
\newcommand{\epic}{\end{tikzpicture}}

\newcommand{\M}{\mathfrak{M}}

\title{A toolkit for ortho-symplectic dualities}

\author[1]{Sergio Benvenuti}
\author[2]{Gabriele Lo Monaco}

\affiliation[1]{Istituto Nazionale di Fisica Nucleare, Sezione di Trieste, Italy}
\affiliation[2]{Institut de Physique Th\'eorique, Universit\'e Paris Saclay, CEA, CNRS, \\ Orme des Merisiers, 91191 Gif-sur-Yvette CEDEX, France}

\emailAdd{benve79@gmail.com}
\emailAdd{gabriele.lomonaco@ipht.fr}

\abstract{We propose new confining dualities in $3d$ $\cN\!=\!2$ gauge theories with orthogonal gauge groups, with and without monopole superpotentials. Deriving some of those dualities requires a sequence of gauging and Higgsing for a $\mathbb{Z}_2$ symmetry.  This  prevents the gauge theory from developing a smooth quantum moduli space and affects the global structure of the gauge group, muting it from $SO$ to $O_+$  . \newline
The confining dualities provide tools to deconfine  fields transforming in the symmetric rank-$2$ representation of classical gauge groups. 
As an application, we propose and derive S-confining dualities for $SO(N)$ $(Sp(N))$ gauge theories with an adjoint and $1$ $(2)$ fundamentals, respectively. From these S-confining dualities, we readily obtain various \emph{duality appetizers} and the $3d$ mirror of $A_{2N}$ Argyres-Douglas theories.}

\begin{document}
\maketitle

\section{Introduction and summary}
One of the most interesting and fascinating aspect of quantum gauge theories at strong coupling is the possibility for two different theories to flow to the same conformal fixed point at long distances. 

The infrared dynamics can sometimes be described only in terms of a few gauge-singlet fields, theories with this property are usually called \emph{S-confining}. In the context of $3d$ $\cN=2$ gauge theories, a paradigmatic example is the duality between SQED with one flavor $(Q,\tilde Q)$ and the XYZ model \cite{Aharony:1997bx}. On the electric side, the infrared degrees of freedom can be understood as the monopole operators $\mathfrak{M}^{\pm}$ and the meson $\mathbb{M}=Q\tilde Q$, mapping to three chiral fields on the magnetic side, interacting through a superpotential of the form $\cW=XYZ$, as the name of the model suggests.

More general S-confining dualities relate gauge theories to Wess-Zumino theories (or free chirals in the particular case of vanishing superpotential). Some of them can be obtained starting from a known Aharony-like duality \cite{Aharony:1997gp} and appropriately choosing the matter content and the superpotential. As an example, $Sp(N-1)$ SQCD with $N_f=2N$ flavors and $\cW=\sigma \M$ ({\it i.e.} flipped monopole superpotential) is dual to a theory of $N(2N-1)$ free chirals, transforming in the rank-two antisymmetric representation of the $SU(2N)$ flavor group. 

Confining dualities are useful to \emph{deconfine} fields transforming in rank-two representations of various gauge groups. This observation has been used in the context of $4d$ $\cN=1$ gauge theories \cite{Berkooz:1995km, Pouliot:1995me, Luty:1996cg,Terning:1997jj} and of $3d$ $\cN=2$ gauge theories \cite{Nii:2016jzi,Pasquetti:2019uop,Pasquetti:2019tix,Benvenuti:2020gvy}. \cite{Benvenuti:2020gvy} \emph{sequentially deconfined} either $U(N)$ gauge theories with an adjoint and $F$ flavors or $Sp(N)$ gauge theories with an antisymmetric and $2F$ fundamentals.

In this paper, we propose new confining dualities of gauge theories with orthogonal gauge groups. We start from a duality for theories with $SO(N)$ gauge groups and matter in the vector representation, found by Aharony, Razamat, Seiberg and Willett (ARSW) \cite{Aharony:2013kma}, building on \cite{Hwang:2011ht,Aharony:2011ci,Kapustin:2011gh,Benini:2011mf}: 
\be
\ba{c} SO(N_c) \, \text{w/ $F$ chiral flavors\,Q} \\ \cW=0 \ea
     \qquad      \Longleftrightarrow \qquad
\ba{c} SO(F-N_c+2) \, \text{w/ $F$ chiral flavors\,q} \\ 
1+F(F+1)/2\,\,\text{singlets}\,\,\sigma, \mathbb{S}^{ij}\\
\cW= \sigma \mathfrak{M}^+ +  \mathbb{S}^{ij} \Tr(q_i q_j)  \ea 
\ee
For $F<N_c-2$, $SO(N_c)$ SQCD has no supersymmetric vacuum, while one may expect confining dualities for $F=N_c-1$ or $F=N_c=2$, which we obtain decoupling flavors in the ARSW duality. In the first case, we propose that the magnetic theory is actually a Wess-Zumino model with an $SU(N_c-1)$ symmetric matrix of chirals $\mathbb{S}$, a vector $q$ and a singlet $\sigma$ with superpotential $\cW=\text{Tr}(q\mathbb{S}q)+\sigma^2 \text{det}\mathbb{S}$, \eqref{eq:SconfW=0}. For $F=N_c-2$, the theory develops a smooth quantum moduli space. However, we propose a mechanism to prevent the moduli space from smoothing: we start from the case with $F=N_c-1$, we turn on a mass term for a flavor, then we gauge a $\mathbb{Z}_2$ symmetry. On the magnetic side, the discrete gauge symmetry is Higgsed while on the electric side, there is no Higgsing and the gauge group is turned from $SO$ to $O_+/\text{Spin}$. We thus arrive to the duality \eqref{eq:QDMSflipped}.

A second starting point is a variation of the ARSW duality with linear monopole superpotential \cite{Aharony:2013kma, Amariti:2018gdc}:
\be
\ba{c} SO(N_c) \, \text{w/ $F$ chiral flavors\,Q} \\ \cW=\mathfrak{M}^+ \ea
     \qquad      \Longleftrightarrow \qquad
\ba{c} SO(F-N_c) \, \text{w/ $F$ chiral flavors\,q} \\ 
\text{and}\,\,\frac{F(F-1)}{2}\,\,\text{singlets} \,\, \mathbb{S}^{ij}\,,\\
\cW= \mathfrak{M}^+ +  \mathbb{S}^{ij} \Tr(q_i q_j)  \ea 
\ee
In this case, there are no supersymmetric vacua for $F<N_c$ and one can expect to find interesting confining dualities for $F=N_c+1$ or $F=N_c$. In the first case, we propose the dual to be a Wess-Zumino model with a symmetric matrix of chirals $\mathbb{S}$, a vector $q$ and superpotential $\cW=\text{Tr}(q\mathbb{S}q)+\text{det}\mathbb{S}$, \eqref{eq:linSO1}. For $F=N_c$ we expect again a smooth quantum moduli space. Building on the previous lesson, we avoid a quantum deformed moduli space taking $SO(N_c)$ SQCD with $N_c+1$ flavors and linear monopole superpotential, turning on a mass term for one flavor and gauging/Higgsing a $\mathbb{Z}_2$ symmetry. On the electric side, the gauged discrete symmetry is preserved and we end up with a confining duality for $O_+(N_c)$ gauge group \eqref{eq:linSOTriangle}.

All such dualities are useful in deconfining rank-two symmetric representations (of $SU(F)$ or subgroups thereof). In particular, we apply such dualities, together with symplectic confining dualities \cite{Aharony:1997gp,Aharony:2013dha}, to find S-confining dualities for $SO(N)$ $(Sp(N))$ gauge theories with an adjoint and $1$ $(2)$ fundamentals, respectively. In both cases, the dual consists of a Wess-Zumino or free chirals depending on the choice of superpotential. Such S-confining dualities can be used in turn as an intermediate step to prove other proposed dualities in the literature as duality appetizers \cite{Jafferis:2011ns,Kapustin:2011vz,Amariti:2014lla,Benvenuti:2018bav}. Turning on a real mass for a flavor, we are able to prove the duality appetizer of \cite{Kapustin:2011vz} between $SO(2N)_1$ with an adjoint $\Phi$ and $N$ free chirals (duals of the composite operators $\text{Tr}(\Phi^{2i}), \text{Pf}\Phi$) and the duality appetizer of \cite{Benvenuti:2018bav} between $Sp(N)_{\frac{1}{2}}$ with and adjoint and one flavor with $N$ free chirals (duals of adjoint traces). Moreover, we propose and derive a new duality appetizer: $SO(2N+1)_1$ with adjoint dual of $N$ free fields, mapping again to the adjoint traces in the electric theory.\footnote{To be precise, the electric theory of the proposed duality has a non-zero superpotential of flipping type. For this reason, this is not actually a proper appetizer in the sense of \cite{Jafferis:2011ns,Kapustin:2011vz,Amariti:2014lla,Benvenuti:2018bav}.} If we keep the CS level to be vanishing but pick an appropriate superpotential, we are able to prove instead that the $3d$ mirror of $(A_1,A_{2N})$ Argyres-Douglas is a theory of $N$ free massless hypermultiplets, as already claimed in \cite{Benvenuti:2018bav}.

In a companion paper \cite{BLM2}, we apply such new confining dualities in order to sequentially deconfine $Sp(N)$, $O_+(2N+1)$ and $SO(2N)$ gauge theories with an adjoint and an arbitrary number of fundamental flavors.

The structure of the paper is the following: in section \ref{sec:Oduals} we review the basic ARSW duality for orthogonal groups and, building on that, we prove new confining dualities with flipped monopole superpotential for orthogonal gauge theories; new confining dualities with monopole superpotential are instead derived in section \ref{sec:Omonduals}. All the dualities useful to deconfine rank-two matter in ortho-symplectic gauge theories are collected in section \ref{sec:handbook}. Those dualities are used in section \ref{sec:SconfSP} to derive S-confining dualities for $SO(N)$ $(Sp(N))$ gauge theories with an adjoint and $1$ $(2)$ fundamentals, respectively. Building on those, we also derive the aforementioned duality appetizers and $3d$ mirror of $(A_1,A_{2N})$ Argyres-Douglas, in section \ref{sec:3dmirrors} and \ref{sec:appetizers}, respectively.

\section{Dualities and confinements for orthogonal gauge theories}\label{sec:Oduals}
In this section we study the $3d$ $\cN=2$ gauge theories with orthogonal groups with $N_c$ colors, and matter in $F$ copies of the vector representation. Building upon the duality first proposed by  Aharony, Razamat, Seiberg and Willett (ARSW) \cite{Aharony:2013kma}, valid for $F \geq N_c$, we discuss the infrared behavior for $F=N_c - 1$ and $F=N_c-2$. In the next section we discuss the same theories with a linear monopole superpotential turned on. A prominent role will be played by discrete global symmetries.

Let us start with a  review of the properties of the discrete symmetries. Different orthogonal groups can differ in their global properties, due to the gauging of discrete global symmetries. The usual $SO(N_c)$ gauge group comes with two global $\mathbb{Z}_2$: the charge-conjugation $\cC$ (or $\mathbb{Z}_2^{\cC}$), that acts on vectors as a reflection and a magnetic symmetry $\cM$ (or $\mathbb{Z}_2^\cM$) associated to non-trivial center of the group and changing sign to the fundamental monopole operator. Gauge-invariant operators in the chiral ring are said baryonic if they are charged under $\cC$ and can be both baryons in the usual terminology (\text{i.e.} composite operators made of chiral fields contracted with the Levi-Civita tensor) or baryon-monopoles, using the terminology of \cite{Aharony:2013kma}. By gauging different combinations of such discrete symmetries, different gauge groups are obtained, usually denoted by $O(N_c)_{\pm}$, $\text{Spin}(N_c)$ and $\text{Pin}(N_c)$. The variouss possible gaugings are summarized in \eqref{eq:gaugingZ2}.
\be
\label{eq:gaugingZ2}
\scalebox{1}{
\begin{tikzpicture}[baseline]
\tikzstyle{every node}=[font=\footnotesize]
\node[draw=none] (node1) at (0,0) {$SO(N_c)$};
\node[draw=none] (node2) at (2.7,1.7) {$O_+(N_c)$};
\node[draw=none] (node3) at (2.7,0) {$\text{Spin}(N_c)$};
\node[draw=none] (node4) at (2.7,-1.7) {$O_-(N_c)$};
\node[draw=none] (node5) at (5.4,0) {$\text{Pin}(N_c)$};
\draw[black,->] (node1) edge node[above]{$\cC$} (node2) ;
\draw[black,->] (node1) edge node[above]{$\cM$} (node3) ;
\draw[black,->] (node1) edge node[below]{$\cC\cM$} (node4) ;
\draw[black,->] (node2) edge node[above]{$\cM$} (node5) ;
\draw[black,->] (node3) edge node[above]{$\cC$} (node5) ;
\draw[black,->] (node4) edge node[right]{$\cM$} (node5) ;
\end{tikzpicture}}
\ee
In general, gauging a discrete symmetry strongly affects the chiral ring, projecting out part of the generators: this is the case, for instance, of the  baryonic operators when charge conjugation $\cC$ is gauged, as in $O(N_c)_+$ and $\text{Pin}(N_c)$. 

\subsection{Dualities}
The duality for orthogonal groups (special and not) proposed in \cite{Aharony:2013kma} generalizes the dualities previously proposed in \cite{Benini:2011mf,Aharony:2011ci,Kapustin:2011gh,Hwang:2011ht} for $O_+(N_c)$ strictly. The starting point is $SO(N_c)$ SQCD with $F$ flavors $Q$. Since the group is special, both $\cC$ and $\cM$ are global symmetries and none of the possible chiral ring generators is projected out. Thus, the global symmetry group of the theory is:
\be
\label{eq:globalSQCD}
G_{\text{glob}}\,=\,SU(F)\times \frac{ U(1)_A\times \mathbb{Z}_2^\cC\times \mathbb{Z}_{2}^\cM}{\mathbb{Z}_2}\,.
\ee
In \eqref{eq:globalSQCD}, $U(1)_A$ denotes the axial symmetry acting on the flavors; moreover, the quotient with respect to a $\mathbb{Z}_2$ factor, is due to the relation $\cC^{N_c}\cdot \cM^{F}\cdot e^{i\pi \mathcal{A}}=1$, where $\mathcal{A}$ the generator of $U(1)_A$. The chiral ring of the theory is generated by the following operators: the meson $\Tr(Q^i\,Q^j)$, transforming in the rank-2 symmetric representation of $SU(F)$ and the baryon $\epsilon_K\!\cdot \!Q^K$ where $\epsilon_K$ denotes the Levi-Civita tensor of $SO(N_c)$ and the contraction of indices is understood. The baryon transforms in the rank-$K$ antisymmetric representation under $SU(F)$ and it has a non-trivial charge under $\cC$. Together with such composite operators we also find monopole operators. In the case of $SO(N_c)$ we have two different kind of monopoles, depending on their charge under $\cC$ action. The unit-flux monopoles are usually denoted as $\mathfrak{M}^{\pm}$ where the sign reveals the charge under $\mathbb{Z}_2^\cC$. The odd monopole $\mathfrak{M}^{-}$, however, it is not invariant on its own, but it actually needs to be dressed with $N_c-2$ flavors;\footnote{In fact, a non trivial vev of a unit-flux monopole breaks the gauge group to $S(O(N_c-2)\times O(2))$ where we stress the presence of a residual gauged $\mathbb{Z}_2$ factor, {\it i.e.} the reflection in both $O(2)$ and $O(N_c-2)$ . The odd monopole is not invariant under such  residual $\mathbb{Z}_2$ factor, and it needs to be dressed with another odd operator (invariant with respect to the continuous part of the residual gauge group); such operator can be built appropriately contracting the Levi-Civita symbol $\epsilon_{N_c-2}$.} the resulting ``baryon-monopole" will be denoted as $\mathfrak{M}^{-}_{\epsilon\cdot Q^{N_c-2}}$, it only exists for $F\geq N_c-2$ and it transforms in the  rank-($N_c-2$) antisymmetric representation of $SU(F)$. Both the even and odd monopole transform non-trivially under the magnetic symmetry $\cM$. All the charges are summarized in table \ref{tab:chargesSQCD}.
\be
\label{tab:chargesSQCD}
\begin{tabular}{ l | c | c | c | c}
      & $SU(F)$ & $U(1)_A$ & $\cC$ & $\cM$\\
      \hline
      \hline
$\Tr(Q^iQ^j)$ & $\text{\bf symm}^{\mathbf{2}}$ & $2$ & $+$ & $+$ \\
$\epsilon_{K}\!\cdot\!Q^K$ & $\text{\bf antisym}^{\mathbf{K}}$ & $K$ & $-$ & $+$ \\
\hline
\hline
$\mathfrak{M}^+ $ & $\text{\bf singlet}$ & $-F$ & $+$ & $-$\\
$\mathfrak{M}^-_{\epsilon\cdot Q^{K-2}} $ & $\text{\bf antisym}^{\mathbf{K-2}}$ & $K-F-2$ & $-$ & $-$\\
\end{tabular}
\ee

$$R[\M^+]=F(1-R[Q])+2-N_c$$

The ARSW duality states that
\be
\label{eq:ARSW}
\ba{c} SO(N_c) \, \text{w/ $F$ chiral flavors\,Q} \\ \cW=0 \ea
     \qquad      \Longleftrightarrow \qquad
\ba{c} SO(F-N_c+2) \, \text{w/ $F$ chiral flavors\,q} \\ 
1+F(F+1)/2\,\,\text{singlets}\,\,\sigma, \mathbb{S}^{ij}\\
\cW= \sigma \mathfrak{M}^+ +  \mathbb{S}^{ij} \Tr(q_i q_j)  \ea 
\ee
  with mapping
\be
\label{eq:ARSWmap}
\left\{ \ba{c} \text{Tr}(QQ) \\ \epsilon \cdot Q^{N_c} \\ \M^+ \\ \M^-_{\epsilon \cdot Q^{N_c-2}}\ea \right\}
     \qquad      \Longleftrightarrow \qquad
\left\{ \ba{c} \mathbb{S} \\ \mathfrak{M}^-_{\epsilon\cdot q^{F-N_c}} \\ \sigma \\ \epsilon \cdot q^{F-N_c+2} \ea \right\} 
   \ee
Duality \eqref{eq:ARSW} is valid for $F \geq N_c$.
As in all Aharony-like dualities, the meson $\Tr(Q^iQ^j)$ in the electric theory maps to the gauge singlet matrix $\mathbb{S}^{ij}$ in the magnetic dual, while the even monopole of the electric theory $\mathfrak{M}^+$ maps to the flipping field $\sigma$ on the magnetic side. The non-trivial part of the ARSW proposal consists in the mapping of the baryonic operators. The baryon $\epsilon_{N_c} \cdot Q^{N_c}$ of the electric theory maps to the baryon-monopole $\mathfrak{M}^-_{\epsilon\cdot q^{F-N_c}}$ of the magnetic dual.\footnote{Observe that all the charges actually match since in the magnetic frame, the chiral fields $q$ have charge $-1$ under $U(1)_A$ and transforms in the anti-fundamental representation of $SU(F)$.} Viceversa, the baryon-monopole of the original theory, $\mathfrak{M}^-_{\epsilon\cdot Q^{N_c-2}} $,  gets mapped to the baryon $\epsilon_{F-N_c+2}\cdot q^{F-N_c+2}$. In order to make the comparison more clear, all the charges of the chiral ring generators of the magnetic theory are collected in table \ref{tab:chargesSQCDmag}.
\be
\label{tab:chargesSQCDmag}
\begin{tabular}{ l | c | c | c | c}
      & $SU(F)$ & $U(1)_A$ & $\cC$ & $\cM$\\
      \hline
      \hline
$\mathbb{S}^{ij}$ & $\text{\bf symm}^{\mathbf{2}}$ & $2$ & $+$ & $+$ \\
$\epsilon_{F-K+2}\cdot q^{F-K+2}$ & $\text{\bf antisym}^{\mathbf{K-2}}$ & $K-F-2$ & $-$ & $+$ \\
\hline
\hline
$\sigma $ & $\text{\bf singlet}$ & $-F$ & $+$ & $-$\\
$\mathfrak{M}^-_{\epsilon\cdot q^{F-K}} $ & $\text{\bf antisym}^{\mathbf{K}}$ & $K$ & $-$ & $-$\\
\end{tabular}
\ee
A careful reader could have observed at this point that a mismatch in the mapping of the $\mathbb{Z}_2$ charges seems to be present: for instance, in the electric frame the baryon is only charged under $\cC_{\text{el.}}$ but its proposed magnetic dual, the baryon-monopole, transforms under both $\cC_{\text{mag.}}$ and $\cM_{\text{mag.}}$. The reason of such mismatch is that there is a non-trivial mapping of the discrete symmetries also:
\be
\label{eq:MapDiscrete}
\cC_{\text{el.}}\\,\,\,\leftrightarrow\,\,\, \cC_{\text{mag.}}\,,\quad \cM_{\text{el.}}\\,\,\,\leftrightarrow\,\,\, (\cC\cdot \cM)_{\text{mag.}}\,.
\ee
The analogue ARSW duality for non-special orthogonal groups are obtained by the appropriate gauging of the discrete symmetries and taking into account \eqref{eq:MapDiscrete}. Let us start with the $O(N_c)_+$ case. Gauging charge conjugation, all the baryonic operators are projected out. The map \eqref{eq:MapDiscrete} implies that in the magnetic theory charge conjugation is also gauged and the two magnetic symmetries maps to each other. Thus the appropriate Aharony-like duality for $O(N_c)_+$ is completely analogous to \eqref{eq:ARSW} with $O_+$ gauge groups in both frame \cite{Benini:2011mf,Aharony:2011ci,Hwang:2011ht}.

The $O(N_c)_-$ theory is less common in literature: in this case $\cC\cdot\cM$ is gauged and thus both the even monopole $\mathfrak{M}^+$ and the baryon $\epsilon_K\cdot Q^K$ are not gauge invariant anymore.\footnote{However, the product $\mathfrak{M}^+(\epsilon_K\!\cdot\!Q^K)$ is still in the chiral ring} The chiral ring still contains the meson $\Tr(Q^iQ^j)$, the baryon monopole and the monopole usually denoted by $\mathfrak{M}^{+}_{\text{Spin}}$, that is the even monopole with two units of magnetic flux. It is evident from the mapping \eqref{eq:MapDiscrete} that if we gauge $\cC_{\text{el.}}\!\cdot\!\cM_{\text{el.}}$ on the electric side, in the magnetic frame we need to gauge $\cM_{\text{mag}.}$, {\it i.e.} the magnetic gauge group $\text{Spin}(F-N_c+2)$. In such a theory, all unit-flux monopoles (either baryonic or not) are not gauge invariant but the chiral ring contains a two-units-flux baryon-monopole (defined analogously to the usual baryon-monopole), together with the two-units-flux even monopole $\mathfrak{M}^{+}_{\text{Spin}}$, the baryon and the meson. The mapping of the operators between the two sides of the duality is straightforward. 

Finally, gauging in a $\text{Pin}(N_c)$ theory all discrete symmetry are gauged and thus all monopoles and baryonic operators are projected out. The chiral ring is generated by the meson and the two-units-flux even monopole. A Pin$(N_c)$ gauge theory is dual to a Pin$(F-N_c+2)$ gauge theory.

\subsection{Confining dualities}
For $F<N_c-2$, the electric theory $SO(N_c)$ with $F$ flavors breaks supersymmetry. The cases $F=N_c-1$ and $F=N_c-2$ are confining, we discuss their low energy behavior below.

\subsubsection{$F=N_c-1$}
If $F=N_c-1$, we propose the following S-confining duality
\be
\label{eq:SconfW=0}
\ba{c} SO(K+1) \, \text{w/ $K$ chiral flavors\,Q} \\ \cW=0 \ea
     \qquad      \Longleftrightarrow \qquad
\ba{c} \text{Wess-Zumino w/\, a singlet $\sigma$,} \\
\text{an $SU(K)$ symmetric $\bS$  and a}\\
\text{$SU(K)$ anti-fundamental $q$\,,}\\
 \cW=q\bS q+\sigma^2\,\text{det}(\bS)\\
  \ea \ee
  with mapping
\be
\label{eq:SconfW=0map}
\left\{ \ba{c} \text{Tr}(QQ) \\ \M^+ \\ \M^-_{\epsilon \cdot Q^{K-1}}\ea \right\}
     \qquad      \Longleftrightarrow \qquad
\left\{ \ba{c} \mathbb{S} \\ \sigma \\ q \ea \right\} 
   \ee
Using that $R[\M^+]=1-K \, R[Q]$, $R[\M^-_{\epsilon\cdot Q^{K-1}}]=1-R[Q]$, it is easy to check that the global symmetries are consistent with the duality and the mapping.\footnote{On the r.h.s. $R[q\bS q]=2R[Q]+2(1-R[Q])= R[\sigma^2\,\text{det}(S)]=2(1-K R[Q])+2 K R[Q] = 2$.}  We also checked duality \eqref{eq:SconfW=0} with the supersymmetric index.

We can derive duality \eqref{eq:SconfW=0} starting from \eqref{eq:ARSW} with $N_c=F=K+1$, so that the r.h.s. is an $SO(2)$ gauge theory with $K\!+\!1$ flavors, $\cW=\sigma \mathfrak{M}^+ +  \bS ^{ij} \Tr(q_i q_j)$. A mass term for the last flavor on the l.h.s. is mapped to $\bS^{K+1,K+1}$, so on the r.h.s. the theory is Higgsed $SO(2) \rightarrow SO(1)$. One of the two components of the original flavors, together with the gauge singlets $\bS^{i, K+1}$ become massive and can be integrated out. Along the flow a superpotential term $\sigma^2\,\text{det}(\bS)$ is consistent with all global symmetries, hence we expect it to be generated by non-perturbative effects.

\subsubsection{$F=N_c-2$}
We now move to the case $F=N_c-2$, which is described in the infrared by a quantum deformed moduli space \cite{Aharony:2013kma}. Starting from \eqref{eq:SconfW=0} with $K \rightarrow K+1$, turning on a mass term of the form $tr(Q^0 Q^0) \leftrightarrow \bS^{00}$, on the r.h.s. the superpotential becomes
\be \cW=\bS^{00} + \bS^{ij} q_i q_j+q_0 S^{i}q_i+ \bS^{00}q^0q^0+\sigma^2\,(\bS^{00}\text{det}(\mathbb{S})+\epsilon^{K}\epsilon_{K}\mathbb{S}^{K-1} S^i S^j)\ee
where we split $\bS \rightarrow \{\mathbb{S}^{ij},S^i,\bS^{00}\}$ and $q \rightarrow \{q_i , q_0\}$, $i,j=1,\ldots,K$. The F-terms of $S^{00}$
\be (q^0)^2 + \sigma^2 \text{det}(\mathbb{S}) + 1 = 0 \ee

Since $q_0$ and/or $\sigma^2 \text{det}(\mathbb{S})$ is taking a non zero vev, $S^i$ and $q_i$ are always massive.  Hence at low energy we have 
\be
\label{eq:QDMSW=0}
\ba{c} SO(K+2) \, \text{w/ $K$ chiral flavors\,Q} \\ \cW=0 \ea
     \qquad      \Longleftrightarrow \qquad
\ba{c} \text{Wess-Zumino w/}\\
\text{ three singlets $\sigma$, $\bS^{00}, q_0$,} \\
\text{an $SU(K)$ symmetric $\bS$\,,}\\
 \cW=\bS^{00}((q^0)^2 + \sigma^2 \text{det}(\mathbb{S}) + 1))\\
  \ea \ee
  with mapping
\be
\label{eq:QDMSW=0map}
\left\{ \ba{c} \text{Tr}(QQ) \\ \M^+ \\ \M^-_{\epsilon \cdot Q^{K}}\ea \right\}
     \qquad      \Longleftrightarrow \qquad
\left\{\ba{c} \mathbb{S} \\ \sigma \\ q_0 \ea\right\}
   \ee

\subsection{Confining dualities with flipped monopole}
Let us begin with $F=N_c-1$,  flipping $\M^+ \leftrightarrow \sigma$ in \eqref{eq:SconfW=0} we obtain the following S-confining duality:
\be
\label{eq:flipSO1}
\ba{c} SO(K+1) \, \text{w/ $K$ chiral flavors\,Q} \\ \cW= \sigma\,\mathfrak{M}^+ \ea
     \qquad      \Longleftrightarrow \qquad
\ba{c} \text{Wess-Zumino w/\,}  \frac{K (K+1)}{2}\,\text{chirals} \\
\text{symmetric of $SU(K)$ $\mathbb{S}^{ij}$}\,,\\
K\,\text{chirals $q_j\,,\mathcal{W}=q_i\,\mathbb{S}^{ij}\,q_j$}
  \ea \ee
Some comments: on the gauge theory side, the chiral ring contains the meson $\Tr(Q^iQ^j)$ and the baryon-monopole $\mathfrak{M}^{-}_{\epsilon_{K-1}\cdot Q^{K-1}}$, the latter transforming in the anti-fundamental representation of flavor group $SU(K)$. For algebraic reasons, it is not possible to build the baryon while the even monopole $\mathfrak{M}^+$ is flipped: thus, we can observe that the discrete symmetry $\cC\!\cdot\!\cM$ acts trivially on the chiral ring of the gauge theory frame. On the Wess-Zumino side, $\mathbb{S}^{ij}$ is the dual of the meson, while $q_i$ is the dual of the baryon monopole. In the magnetic frame, we can still introduce a charge-conjugation $\cC$, charging $q$ and coinciding with three-dimensional reflection; we do not have, instead, any magnetic discrete symmetry, consistently with the fact that l.h.s. $\cC\!\cdot\!\cM$ has a trivial action on the chiral ring. Gauging any combination of the discrete symmetries in l.h.s. of \eqref{eq:flipSO1}, the duality is not confining anymore: the magnetic theory would be an $O(1)_+$ gauge theory.

Duality \eqref{eq:flipSO1} can be useful to deconfine a symmetric field coupled to a flavor, we will use it in section \ref{sec:Sconf} and in \cite{BLM2}.

We can study the case $F\!=\!N_c-2$ either turning on a mass term for a single flavor in \eqref{eq:flipSO1},  or flipping $\M^+ \leftrightarrow \sigma$ in \eqref{eq:QDMSW=0}. On the r.h.s. we obtain a quantum deformed moduli space with the equation
$(q_0)^2 = -1$.
So there are two disconnected branches. Recall that $q_0$ maps to the baryon monopole $\M^-_{\epsilon \cdot Q^{K}}$ on the gauge theory side.

This means that  we can do the following: we gauge $\cM$ or $\cC$ in the gauge theory side, in the r.h.s. we then have a $\mathbb{Z}_2$ gauge theory, which is however Higgsed down to a trivial gauge group going on one of the two branches $q_0 = \pm i$.
We then obtain the following duality 
\be
\label{eq:QDMSflipped}
\ba{c} O_+(K+2)\, \text{or} \,Spin(K+2)\\
     \text{w/ $K$ chiral flavors\,Q} \\ 
     \cW= \sigma\,\mathfrak{M}^+ \ea
     \qquad      \Longleftrightarrow \qquad
\ba{c} \text{Free $SU(K)$ symmetric} \\
    \bS \leftrightarrow \text{Tr}(QQ) \ea 
  \ee
Duality \eqref{eq:QDMSflipped} can be used to deconfine any symmetric field, even if we do not use it, neither in this paper nor in \cite{BLM2}.

\section{Dualities and confinement for orthogonal gauge theories with monopole superpotential}\label{sec:Omonduals}

\subsection{Dualities}
The following \emph{monopole duality} for $SO$ gauge groups is valid for $F \geq N_c+2$ \cite{Aharony:2013kma,Amariti:2018gdc}:
\be
\label{eq:ARSWlinear}
\ba{c} SO(N_c) \, \text{w/ $F$ chiral flavors\,Q} \\ \cW=\mathfrak{M}^+ \ea
     \qquad      \Longleftrightarrow \qquad
\ba{c} SO(F-N_c) \, \text{w/ $F$ chiral flavors\,q} \\ 
\text{and}\,\,\frac{F(F-1)}{2}\,\,\text{singlets} \,\, \mathbb{S}^{ij}\,,\\
\cW= \mathfrak{M}^+ +  \mathbb{S}^{ij} \Tr(q_i q_j)  \ea 
\ee
  with mapping
\be
\label{eq:ARSWlinearmap}
\left\{ \ba{c} \text{Tr}(QQ) \\ \epsilon \cdot Q^{N_c} \ea \right\}
     \qquad      \Longleftrightarrow \qquad
\left\{ \ba{c} \mathbb{S}  \\ \epsilon \cdot q^{F-N_c} \ea \right\} 
   \ee
On both sides, the linear monopole superpotential breaks the magnetic symmetry $\cM$ while charge conjugation is preserved. The linear monopole superpotential also breaks the $U(1)$ axial symmetry and the R-charge of the chiral fields is fixed to be 
\be R[Q]=1-\frac{N_c}{F} \,, \qquad \qquad R[q]=1-\frac{F-N_c}{F}\,.\ee
 The chiral ring is generated by the meson and the baryon only.

\eqref{eq:ARSWlinear} is related to \eqref{eq:ARSW}: starting from \eqref{eq:ARSW} with two additional flavors, when we turn on $\M^+ \leftrightarrow \sigma$, on the r.h.s. the monopole takes a vev, forced by the F-terms of $\sigma$. Such a vev implies a Higgsing $SO(F+2-N_c) \rightarrow SO(F-N_c)$. Along the RG flow the linear monopole term $\M^+$, preserving all global symmetries, is generated.

 If $F<N_c$ a runaway superpotential is generated and supersymmetry is broken. The cases $F=N_c+1$ and $F=N_c$ are instead confining, and we analyze them below.

\subsection{Confining dualities}
In the case $F=N_c+1$ we propose the following S-confining duality:
\be\label{eq:linSO1}
\ba{c} SO(K-1) \, \text{w/ $K$ chiral flavors\,Q} \\ \cW= \mathfrak{M}^+ \ea
     \qquad      \Longleftrightarrow \qquad
\ba{c} \text{Wess-Zumino w/\,}   \\
\text{$SU(K)$ symmetric $\,\,\mathbb{S}^{ij}$},\\
SU(K)\, \text{antifundamental} \,q_i\,, \\
\mathcal{W}=q_i\,\mathbb{S}^{ij}\,q_j + \text{det}(\mathbb{S})
  \ea \ee
    with mapping
\be
\label{eq:ARSWlinearmap}
\left\{ \ba{c} \text{Tr}(QQ) \\ \epsilon \cdot Q^{K-1} \ea \right\}
     \qquad      \Longleftrightarrow \qquad
\left\{ \ba{c} \mathbb{S}  \\ q \ea \right\} 
   \ee
As a check of duality \eqref{eq:linSO1}, we notice that the marginality of the superpotential implies, on the l.h.s., $R[Q]=\frac{1}{K}$, on the r.h.s. $R[\mathbb{S}]=\frac{2}{K}, R[q_i]=\frac{K-1}{K}$, consistently with the mapping stated above. Moreover, we notice that in the special case $K=3$, duality \eqref{eq:linSO1} coincides with a known duality for $U(1)$ gauge group.\footnote{In the special case $K=3$, the l.h.s. is equal to $U(1)$ with $(3,3)$ flavors and $\cW=\M^++\M^-$, which is known to satisfy the following confining duality \cite{Benvenuti:2016wet}:
\be\label{U1duality}
\ba{c} U(1) \, \text{w/ $(3,3)$  flavors\,$Q,\tilde{Q}$} \\ \cW= \mathfrak{M}^{+}+\M^- \ea
     \qquad      \Longleftrightarrow \qquad
\ba{c} \text{Wess-Zumino w/ a }  3 \times 3 \\
\text{matrix of chirals $\mu$}\,,\\
\mathcal{W}=\text{det}_{3 \times 3}(\mu)\,.
  \ea \ee
The global symmetry is $SU(3) \times SU(3)$, under which the matrix $\mu$ transforms as a bifundamental. Under the diagonal subgroup $SU(3)$, $\mu$ splits into a symmetric $S$ and and antisymmetric $A$, defining $q^i=\epsilon^{ijk}A_{jk}$, $\text{det}(\mu) =  \mathbb{S}^{ij} \Tr(q_i q_j) + \text{det}(\mathbb{S})$, in agreement with \eqref{eq:linSO1}.} 
The duality \eqref{eq:linSO1} can be tested using the superconformal index. For electric theory in \eqref{eq:linSO1} the supersymmetric index reads:
\begin{equation}
\label{eq:IndexEl1}
\mathcal{I}_{\text{el.},\eqref{eq:linSO1}}(x,\boldsymbol{\mu})=\sum_{\mathbf{m}}\frac{1}{|\mathcal{W}|_{\mathbf{m}}}\int\frac{\text{d}\mathbf{z}}{2\pi i \mathbf{z}}Z^{SO(K)}_{\text{vec}}(\mathbf{z}) Z_{\text{mat}}(x,\mathbf{z}_{\text{\bf fund}},\boldsymbol{\mu}_{\text{\bf fund}})\,,
\end{equation}
where the functions appearing in the previous expressions are reviewed in appendix \ref{app:index}. We denoted by $\mathbf{m},\mathbf{z}$ the units of GNO magnetic fluxes and the fugacities of the gauge group respectively. $\boldsymbol{\mu}$ denotes instead the fugacities of $SO(K+1)$ flavor group and $x$ is the R-charge fugacity. In the matter contributions, we explicitly stressed how the fugacities organize.  Observe that in \eqref{eq:linSO1}, the magnetic symmetry of the orthogonal gauge group is explicitly broken by the linear monopole in the superpotential; moreover, for $K$ even, the most interesting case for us in the following, the charge conjugation can actually be identified with an element of the flavor group, as reviewed below \eqref{eq:globalSQCD}.
The index \eqref{eq:IndexEl1} can be evaluated expanding the integrand as a power series in the R-charge fugacity $x$. Evaluating it for various choices of $K$ and flavors R-charge, one always get:
\begin{equation}
\mathcal{I}_{\text{el.},\eqref{eq:linSO1}}=1+\chi_{\text{\bf sym}^{\mathbf 2}}(\boldsymbol{\mu})x^{\frac{2}{K}}+\chi_{\overline{\text{\bf fund}}}(\boldsymbol{\mu})x^{1-\frac{1}{K}}+\dots
\end{equation}
Here $\chi_{\mathcal{R}}(\boldsymbol{\mu})$ stands for the character of the representation $\mathcal{R}$ of the flavor symmetry group $SU(K)$.
One can easily check that this expansion matches the one for the magnetic theory of \eqref{eq:linSO1}.

Duality \eqref{eq:linSO1} can be obtained deforming duality \eqref{eq:ARSWlinear} with $N_c=K-1$, $F=K+1$, turning on a mass term for a single flavor. On the r.h.s. the gauge group $SO(2)$ is Higgsed and the superpotential term $\text{det}(\mathbb{S})$ is generated. 
Alternatively, we can obtain duality \eqref{eq:linSO1} from  \eqref{eq:ARSW} with $N_c=K+1$, $F=K$, so that the r.h.s. is $SO(2)$ with $K$ flavors, $\cW=\sigma \mathfrak{M}^+ +  \mathbb{S}^{ij} \Tr(q_i q_j)$. Turning on $\M^+ \leftrightarrow \sigma$, on the r.h.s. the F-terms of $\sigma$ force the monopole to take a non zero vev, so the gauge group $SO(2)$ is Higgsed.


We now move to the case of $SO(N_c)$ with $N_c$ flavors. If we simply add a mass term to  \eqref{eq:linSO1} we obtain a duality with a quantum deformed moduli space on the r.h.s. Since in this paper we are interested in confining dualities useful to deconfine symmetric matter, we flip operators in the gauge theory side so that the quantum deformed moduli space is replaced by a free theory. As a first step, let us flip  \eqref{eq:linSO1} (with $K \rightarrow K+2$) as
\be\label{eq:linSO1rw}
 \scalebox{1}{
\bpic  
\path (-3,-1) node[red](x1) {$SO(\!K\!+\!1\!)$} -- (-0.5,-1) node[rectangle,draw](x2) {\small{$K$}} -- (-3,-2.3) node[rectangle,draw](x3) {$1$}-- (-3,0.6) node[rectangle,draw](x4) {$1$};
\draw [-] (x1) to (x2); 
\draw [-] (x1) to (x4); 
\draw [-] (x1) to (x3);
\draw [-] (x2) to (x3); 
\node at (-2.5,-3.2){$ \CW=\mathfrak{M}^++u\Tr(Q Q_{K\!+\!1})+\alpha \Tr(Q_{K\!+\!1} Q_{K\!+\!1})$};
\node[below left] at (-2.9,+0.2){$Q_{0}$};
\node[below left] at (-2.9,-1.4){$Q_{K\!+\!1}$};
\node[below right] at (-1.7,-1.6){$u$};
\node[below right] at (-3.3,-1.35){$\times$};
\node at (-1.4,-0.7){$Q$}; 
\node at (1.9,-1){$ \Longleftrightarrow$};
\node at (6.,0.6){WZ model with};
\node at (6.,-0.2){$S^{ij},S^{i\, 0},S^{00}, S^{0\,K\!+\!1}$ (from $\mathbb{S}$)};
\node at (6.,-0.8){$q_{K\!+\!1}, q_i, q_0$ (from $q$)};
\node at (6.,-2.2){$\cW=q_iS^{ij}q_j+ S^{00} q_0^2+ S^{0\,K\!+\!1} q_{0} q_{K\!+\!1}+$};
\node at (6.,-2.8){$+ S^{i\, 0} q_i q_0+ (S^{0\,K\!+\!1})^2\, \text{det}(S^{ij})$};
\epic}
 \ee
 where $i,j=1,2,\ldots,K$. Now we deform this duality with $\delta \cW = \Tr(Q_{0}Q_{0}) \leftrightarrow S^{00}$. The F-terms of $S^{00}$, $q_0^2=-1$, imply that $q_0$ (which maps into a baryon monopole) takes a vev. The vev of $q_0$ in turn makes the fields $S^{0\,K\!+\!1}, q_{K\!+\!1}, S^{i\, 0}, q_i$ massive. Only the $SU(K)$-symmetric chiral field $S^{ij}$ remains massless. There are two disconnected branches, so the r.h.s. becomes the direct sum of two copies of free $SU(K)$-symmetric chiral fields (in the sense that the Hilbert space is the direct sum of the Hilbert space of one copy of of free $SU(K)$-symmetric chiral field). 
 
 In order to obtain a well defined theory on the r.h.s. we gauge both sides of the duality  the $\mathbb{Z}_2$ global symmetry under which $\epsilon\!\cdot\!(Q^K Q_{K+1}) \leftrightarrow q_0$ is charged. This means that on the r.h.s. the two branches are gauge equivalent, so the $\mathbb{Z}_2$ is Higgsed and we are left with a single copy of a free theory. On the l.h.s. the $\mathbb{Z}_2$ global symmetry is charge conjugation $\mathcal{C}$, so the gauge group $SO(K+1)$ becomes $O_+(K+1)$. We thus obtain the following duality:
\be
\label{eq:linSOTriangle}
 \scalebox{1}{
\bpic  
\path (-3,-1) node[red](x1) {$O_+(\!K\!+\!1\!)$} -- (-0.5,-1) node[rectangle,draw](x2) {\small{$K$}} -- (-3,-2.3) node[rectangle,draw](x3) {$1$};
\draw [-] (x1) to (x2); 
\draw [-] (x1) to (x3);
\draw [-] (x2) to (x3); 
\node at (-2.5,-3.2){$ \CW=\mathfrak{M}^++u\Tr(Qv)+\alpha \Tr(vv)$};
\node[below right] at (-2.9,-1.4){$v$};
\node[below right] at (-1.7,-1.6){$u$};
\node[below right] at (-3.3,-1.35){$\times$};
\node at (-1.4,-0.7){$Q$}; 
\node at (1.9,-1.7){$ \Longleftrightarrow$};
\node at (5.5,-1.5){Free $SU(K)$-symmetric};
\node at (5.5,-2.2){$\text{chirals} \,\, \mathbb{S} \leftrightarrow \Tr(QQ)$};
\epic}
 \ee
This duality will be of crucial relevance in the sequential deconfining of ortho-symplectic theories with adjoint in \cite{BLM2}. On the gauge theory side, the linear monopole superpotential explicitly breaks the magnetic discrete symmetry $\cM$ and an Abelian symmetry, so that the global symmetry is $SU(K) \times U(1)_Q$, where the axial $U(1)_Q$ charges $+1$ the flavors $Q$. The superpotential fixes the R-charges of the other fields:
\be
R[v]\,=\,-K\,R_Q\,,\quad R[u]=2+(K\!-\!1)R_Q\,,\quad R[\alpha]=2+2 K\,R_Q\,.
\ee
Let us observe that the field $v$ has an R-charge below $\frac{1}{2}$, but this is not actually a problem, since $\Tr(vv)$ is flipped, so it is zero in the chiral ring, and the are no chiral ring operators below the unitarity bound. The singlets $u^i$ are needed in order to flip the meson $\Tr(vQ)$ that otherwise would be part of the chiral ring, spoiling the duality. The unique generator is actually the meson $\Tr(Q^iQ^j)$, mapping to free symmetric chiral fields $\mathbb{S}^{ij}$ on the Wess-Zumino side.  A crucial test for the duality \eqref{eq:linSOTriangle} is provided by the computation of the supersymmetric index. This can be performed in the same fashion as for the duality \eqref{eq:linSO1}:
\begin{equation}
\label{eq:IndexEl2}
\mathcal{I}_{\text{el.},\eqref{eq:linSOTriangle}}=1+\phi^2 \chi_{\text{\bf sym}^{\boldsymbol{2}}}(\boldsymbol{\mu})\,x^{2R_Q}+\phi^4\left(\chi_{\text{\bf sym}^{\boldsymbol{4}}}+\chi_{\text{\bf box}_{\boldsymbol{2\times 2}}}\right)x^{4R_Q}+\dots
\end{equation}
where $\text{\bf box}_{\boldsymbol{2\times 2}}$ is the $(0,2,0,\dots,0)$ representation of $SU(K)$, a $2\times 2$ box in terms of Young tableaux. $\phi$ denotes the fugacity of the $U(1)_Q$ axial symmetry.
The index \eqref{eq:IndexEl2} coincides with the index of $K(K+1)/2$ free chirals of R-charge $2 R_Q$ and transforming in the rank-2 symmetric representation of $SU(K)$.

\section{A hand-book of $3d$ $\cN=2$ (de)confining ortho-symplectic dualities}\label{sec:handbook}
In this short chapter we summarize the dualities that are useful to confine fields transforming in the symmetric and antisymmetric representation and/or to confine symplectic or orthogonal gauge groups. We will use these dualities in Section \ref{sec:Sconf}.

For antisymmetric fields / symplectic gauge groups, the following two dualities are useful:
\be
\label{eq:decantisymmSU2N}
 \scalebox{1}{
\bpic  
\path (-3,-1) node[blue](x1) {$Sp(\!N-1\!)$} -- (-0.5,-1) node[rectangle,draw](x2) {\small{$2N$}} ;
\draw [-] (x1) to (x2); 
\node at (-2.5,-2){$ \CW= \gamma \, \mathfrak{M}$};
\node at (-1.6,-0.7){$Q$}; 
\node at (1.9,-1.7){$ \Longleftrightarrow$};
\node at (6.,-1.8){Free $SU(2N)$-antisymmetric};
\epic}
\ee
(this is a variation of Aharony duality.\footnote{\label{foot:Pfaffian} Let us start from the original duality proposed in \cite{Aharony:1997gp}:
\be\label{AharonySconf}
\ba{c} Sp(N-1) \, \text{w/ $2N$ chiral flavors} \\ \cW= 0 \ea
     \qquad      \Leftrightarrow \qquad
\ba{c}   \text{Wess-Zumino w/} \, 2N \times 2N  \, \text{antisymmetric  }\\ \text{matrix of chiral fields $A$, and a singlet $\sigma$ }\\ \cW= \sigma\, \text{Pfaff}(A)  \ea \ee
In this duality the monopole is mapped to $\sigma$ ($\mathfrak{M} \leftrightarrow \sigma$), so if we flip the monopole in the l.h.s. with a gauge singlet $\gamma$, on the r.h.s. we obtain a superpotential term $\sigma \gamma$, so $\sigma$ and $\gamma$ become massive, integrating them out the superpotential becomes zero and we obtain the duality  \eqref{eq:decantisymmSU2N}.})
\be
\label{eq:decantisymmSU2N+1}
 \scalebox{1}{
\bpic  
\path (-3,-1) node[blue](x1) {$Sp(\!N-1\!)$} -- (-0.5,-1) node[rectangle,draw](x2) {\small{$2N\!+\!1$}} -- (-3,-2.3) node[rectangle,draw](x3) {$1$};
\draw [-] (x1) to (x2); 
\draw [-] (x1) to (x3);
\draw [-] (x2) to (x3); 
\node at (-2.5,-3.2){$ \CW=\mathfrak{M}+u\Tr(Qv)$};
\node[below right] at (-2.9,-1.4){$v$};
\node[below right] at (-1.7,-1.6){$u$};
\node at (-1.6,-0.7){$Q$}; 
\node at (1.9,-1.7){$ \Longleftrightarrow$};
\node at (6.,-1.8){Free $SU(2N\!+\!1)$-antisymmetric};
\epic}
\ee
(this is a variation of $4d$ Intriligator-Pouliot S-confining duality, reduced to $3d$.)

For symmetric fields / orthogonal gauge groups, the following two dualities, found in the previous sections, are useful:

\be
\label{eq:decsymm}
 \scalebox{1}{
\bpic  
\path (-3,-1) node[red](x1) {$O_+(\!N\!+\!1\!)$} -- (-0.5,-1) node[rectangle,draw](x2) {\small{$N$}} -- (-3,-2.3) node[rectangle,draw](x3) {$1$};
\draw [-] (x1) to (x2); 
\draw [-] (x1) to (x3);
\draw [-] (x2) to (x3); 
\node at (-2.5,-3.2){$ \CW=\mathfrak{M}^++u\Tr(Qv)+\alpha \Tr(vv)$};
\node[below right] at (-2.9,-1.4){$v$};
\node[below right] at (-1.7,-1.6){$u$};
\node[below right] at (-3.3,-1.35){$\times$};
\node at (-1.4,-0.7){$Q$}; 
\node at (1.9,-1.7){$ \Longleftrightarrow$};
\node at (5,-1.8){Free $SU(N)$-symmetric};
\epic}
 \ee

\be
\label{eq:decsymmplusflav}
 \scalebox{1}{
\bpic  
\path (-3,-1) node[red](x1) {$SO(\!N\!+\!1\!)$} -- (-0.5,-1) node[rectangle,draw](x2) {\small{$N$}};
\draw [-] (x1) to (x2); 
\node at (-2.5,-2.2){$ \CW=\gamma \, \mathfrak{M}^+$};
\node at (-1.4,-0.7){$Q$}; 
\node at (1.9,-1.7){$ \Longleftrightarrow$};
\node at (6.,-0.9){Wess-Zumino w/\, $SU(N)$-symmetric $S^{ij}$};
\node at (6.,-1.7){and $SU(N)$-antifundamental $q_i$,};
\node at (6.,-2.5){$ \mathcal{W}=q_i\,S^{ij}\,q_j$};
\epic}
 \ee

\section{S-confining dualities for $SO/Sp$ with adjoint}\label{sec:Sconf}
In this section we apply the deconfining dualities proposed in sections \ref{sec:Oduals} and \ref{sec:Omonduals}, summarized in section \ref{sec:handbook}, in order to find examples of S-confining $3d$ $\cN=2$ theories. 

We show that $Sp(N)$ with adjoint $\Phi$ and two fundamentals $p,q$, with cubic superpotential $\cW= \Phi p p$, is S-confining:
\be
\label{eq:Sconf0}
\ba{c} Sp(N) \, \text{w/ adjoint $\Phi$}\\
   \text{and $2$ fundamentals\,$q$, $p$} \\ 
   \cW= \Phi p p \ea
     \qquad      \Longleftrightarrow \qquad
\ba{c} \text{Wess-Zumino w/\,}  4N \,\text{chirals} \\
 \mathfrak{m}_j \leftrightarrow \M_{\Phi^j},\, j=0,\ldots,2N-1 \\
 \mu_i \leftrightarrow \text{Tr}(\Phi^{2i+1} qq),\, i=0,\dots,N-1\,,  \\
 \sigma_k \leftrightarrow \text{Tr}(\Phi^{2k}),\, k=1,\dots,N\,,  \\
  \ea \ee
 
Similarly, we show that $SO(K)$ with adjoint $\Phi$ and one fundamentals $t$, with vanishing superpotential, is S-confining, either if $K=2N$:
\be
\label{eq:Sconf2}
\ba{c} SO(2N) \, \text{w/ adjoint $\Phi$}\\
   \text{and $1$ fundamental\,$t$} \\ 
   \cW= 0 \ea
     \qquad      \Longleftrightarrow \qquad
\ba{c} \text{Wess-Zumino w/\,}  4N \,\text{chirals} \\
 \mathfrak{m}_j \leftrightarrow \M_{\Phi^j},\, j=0,\ldots,2N-2\\
 \mu_i \leftrightarrow \text{Tr}(t \Phi^{2i} t),\, i=0,\ldots,N-1\\
 \sigma_k \leftrightarrow \text{Tr}(\Phi^{2k}),\, k=1,\ldots,N-1\\
 \mathfrak{m}_{-} \leftrightarrow \text{Baryon-monopole}\,\,\M^-_{\Phi^{N-1}}\\
 \mathbb{B} \leftrightarrow \text{Baryon}\,\,\text{Pf}\,\Phi\,.
  \ea \ee
or $K=2N+1$, in which case:
\be
\label{eq:Sconf3}
\ba{c} SO(2N+1) \, \text{w/ adjoint $\Phi$}\\
   \text{and $1$ fundamental\,$t$} \\ 
   \cW= 0 \ea
     \qquad      \Longleftrightarrow \qquad
\ba{c} \text{Wess-Zumino w/\,}  4N+2 \,\text{chirals} \\
 \mathfrak{m}_j \leftrightarrow \M_{\Phi^j},\, j=0,\ldots,2N-1\\
 \mu_i \leftrightarrow \text{Tr}(t \Phi^{2i} t),\, i=0,\ldots,N-1\\
 \sigma_k \leftrightarrow \text{Tr}(\Phi^{2k}),\, k=1,\ldots,N\\
 \mathfrak{m}_{-} \leftrightarrow \text{Baryon-monopole}\,\,\M^-_{t\Phi^{N-1}}\\
 \mathbb{B} \leftrightarrow \text{Baryon}\,\,\epsilon\!\cdot\!(t\,\Phi^{N})\,.
  \ea \ee

The superpotential on the r.h.s. of \eqref{eq:Sconf0},\eqref{eq:Sconf2} and \eqref{eq:Sconf3} is a complicated polynomial in the massless fields. The superpotential simplifies if we flip the powers of the adjoint $\text{Tr}(\Phi^{2k})$ and, in the $SO$ case, the baryon and the baryon-monopole. In this way the most general superpotential consistent with the $U(1)\times U(1) \times U(1)_R$ global symmetry\footnote{In the electric theories, the two non-R U(1) factors act on $\Phi$ and $t$ respectively.} is a cubic polynomial which we write explicitly in eqs. \eqref{eq:SconfSPflip}, \eqref{eq:SconfSOeflip} and \eqref{eq:SconfSOoflip}.
In this way the duality is also unitary: all the gauge invariant operators are above the unitarity bound $R > \frac{1}{2}$.

From these S-confining dualities, it is easy to prove that the $3d$ reduction of the $4d$ $\cN=1$ Lagrangian for $(A_1, A_{2N})$ Argyres-Douglas theory is dual to free hypers, as proposed in \cite{Benvenuti:2018bav}, see next section. Moreover from  \eqref{eq:Sconf0}, turning on real masses for the flavors, one can immediately obtain the \emph{duality appetizers}.

All the dualities derived in this section applying the confining ones of section \ref{sec:Oduals}, can be further checked using the supersymmetric index, in the same way as we did for the duality \eqref{eq:linSO1}. 

\subsection{$Sp(N)$ with adjoint and $2$ fundamentals}\label{sec:SconfSP}
We now prove the above S-confining dualities using a sequential deconfinement strategy, as in \cite{Pasquetti:2019uop, Pasquetti:2019tix, Benvenuti:2020gvy, BB1}. We start with the theory on the l.h.s. of  \eqref{eq:Sconf0} in quiver notation:
\be\label{T1}
 \scalebox{0.9}{
\bpic  
\path (-5,0) node{$\CT_1:$} -- (-3,0) node[blue](x1) {$C_N$} -- (-3,-1.3) node[rectangle,draw](x2) {$1$}-- (-3+1.3,0) node[rectangle,draw](x3){$1$};
\draw[-] (x1) edge [out=55,in=125,loop,looseness=3] node[above]  {$\Phi$}(x1);
\node at (1,-0.3){$ \cW=\Tr(p\Phi p)$};
\draw[-] (x1) edge node[right]{$q$} (x2);
\draw[-] (x1) edge node[above]{$p$} (x3);
 \epic }
 \ee
 The continuos global symmetry is $U(1)_{\Phi} \times U(1)_{q} \times U(1)_R$. We denote the R-charge of the adjoint field $\Phi$ by $r_\Phi$ and the R-charge of the chiral $q$ by $r_q$; it is useful to remember that:
\be\label{chargesSconf}
R[p]\,=\,1-\frac{r_\Phi}{2}\,,\quad R[\mathfrak{M}]\,=\,1-r_q-(2N-1/2)r_\Phi\,.
\ee

One could think of deconfining the symplectic adjoint field using the duality \eqref{eq:linSOTriangle}. However, given the particular form of the superpotential in \eqref{T1}, with the  cubic superpotential $\Phi p p$, it is more convenient  to deconfine using the duality \eqref{eq:flipSO1}. The resulting dual theory is thus a two-node quiver:
\be
\scalebox{1}{
\begin{tikzpicture}[baseline]
\tikzstyle{every node}=[font=\footnotesize]
\node[draw=none, red] (node1) at (0,0) {$B_{N}$};
\node[draw=none, blue] (node2) at (2,0) {$C_N$};
\node[draw, rectangle] (sqnode) at (2,-1.2) {$1$};
\draw[black,solid] (node1) edge node[below]{$b$} (node2);
\draw[black,solid] (node2) edge node[right]{$q$} (sqnode);
\node at (4,-0.3){$ \cW=\gamma\,\mathfrak{M}^{+,o}$};
\node[draw=none] at (-2,0) {$\mathcal T_{1'}:$};
\end{tikzpicture}}
\ee
Observe that:
\be
R[b]=\frac{r_{\Phi}}{2}\,,\quad R[\mathfrak{M}^{+,o}]\,=\,1-N r_\Phi\,\,\Rightarrow\,\,R[\gamma]\,=\,1+N r_\Phi\,.
\ee
Now we can dualize the symplectic node in the quiver with duality \eqref{AharonySconf}.  Seiberg-like dualities in $3d$ $\cN=2$ quivers, with particular emphasis on monopole operators, were studied in \cite{Benvenuti:2020wpc}, see also \cite{Okazaki:2021gkk}. The symplectic node confines, so we get an $SO(2N+1)$ theory and flow to the following dual frame:
\be 
 \scalebox{0.9}{
\bpic  
\path (-5,0) node{$\CT_2:$} -- (-3,0) node[red](x1) {$B_N$} -- (-3,-1.3) node[rectangle,draw](x2) {$1$};
\draw[-] (x1) edge [out=55,in=125,loop,looseness=3] node[above]  {$A$}(x1);
\node at (1.5,-0.6){$ \cW=\gamma \M^{-}_{\epsilon \cdot A^{N-1}t}+ \sigma^A \epsilon_{2N+1}(A^{N}t) $};
\draw[-] (x1) edge node[right]{$t$} (x2);
 \epic }
 \ee
The superpotential of the theory $\cT_2$ deserves some comments. Since the symplectic node confined, the linear monopole superpotential  is lifted.\footnote{Had the symplectic group not confined, the superpotential would contain flipped monopoles with one unit of magnetic flux for the symplectic node turned on.} However, we still see the flipping field $\gamma$  in the superpotential, flipping the baryon monopole $ \M^{-}_{\epsilon \cdot A^{N-1}t}$. Such superpotential term is dynamically generated; indeed $\gamma$ has the appropriate R-charge (and more in general, quantum numbers) to flip the baryon monopole:
\be
R[ \M^{-}_{\epsilon \cdot A^{N-1}t}]\,=\,R[\mathfrak{M}^-]+(N-1)\,r_\Phi+r_t\,=\,1-N r_\Phi\quad \Rightarrow\quad R[\gamma]+R[\M^{-}_{\epsilon \cdot A^{N-1}t}]=2\,.
\ee
The superpotential term flipping the baryon instead is a direct consequence of the confinement of the symplectic node. In fact, as stressed in the footnote \ref{foot:Pfaffian}, when the symplectic node confines, this generates a Pfaffian term and indeed one can rewrite:
\be
\epsilon_{2N+1}(A^{N}t)\,=\,\text{Pfaff}\left(\begin{matrix}A &-t\, \\ +t\, & 0\end{matrix}\right)\,.
\ee
Now we proceed to deconfine the adjoint field in $\cT_2$ using the deconfinement duality with linear monopole superpotential \eqref{eq:decantisymmSU2N+1}:
\be
\scalebox{1}{
\begin{tikzpicture}[baseline]
\tikzstyle{every node}=[font=\footnotesize]
\node[draw=none, blue] (node1) at (0,0) {$C_{N-1}$};
\node[draw=none, red] (node2) at (2,0) {$B_N$};
\node[draw, rectangle] (sqnode) at (2,-1.2) {$1$};
\node[draw, rectangle] (auxnode) at (1,1.2) {$1$};
\draw[black,solid] (node1) edge node[below]{$c$} (node2);
\draw[black,solid] (node2) edge node[right]{$t$} (sqnode);
\draw[black,solid] (node2) edge node[right]{$u$} (auxnode);
\draw[black,solid] (node1) edge node[left]{$v$} (auxnode);
\node at (1.3,-2.3){$ \cW=\mathfrak{M}^{\bullet,o}+\sigma^A\,\Tr(ut)+\gamma\M^{0,-}_{\epsilon \cdot c^{2N-2}t}+\Tr(vcu)$};
\node[draw=none] at (-2,0) {$\mathcal T_{2'}:$};
\end{tikzpicture}}
\ee
In $\cT_{2'}$, the baryon monopole have the same definition as before, supplemented by the prescription $A\rightarrow (cc)$ while the baryon has been replaced by $\Tr(ut)$. The R-charges of the auxiliary fields $u$ and $v$ are:
\be
R[v]=2-(2N\!+\!1)\frac{r_\Phi}{2}\,,\quad R[u]=Nr_\Phi\,.
\ee
We can now dualize the orthogonal node: also in this case, the node confines and we are left with a symplectic gauge theory. We use duality \eqref{eq:SconfW=0} flipped by $v$, $\sigma^A$ and $\gamma$:
\be
\label{eq:SconfW=01}
\ba{c} SO(2N+1) \, \text{w/}\\
 (2N-2)_c+1_u+1_t \,\, \text{chiral flavors} \\
    \cW= v\, \text{Tr}(u c)  + \sigma^A\,\Tr(ut) + \gamma\M^{-}_{\epsilon \cdot c^{2N-2}t} \ea
     \quad      \Longleftrightarrow \quad
\ba{c} \text{Wess-Zumino w/\, } \\
\{S, \beta,\mu_0,\tilde{q} \} \leftrightarrow \{cc, uu, tt, ct\}\\
\{\tilde{p}, \mathfrak{m}_{2N-1}, \mathfrak{m}_0\}  \leftrightarrow   \{\M^{-}_{\epsilon \cdot c^{2N-3}ut} , \M^{-}_{\epsilon \cdot c^{2N-2}u}, \M^+\} \\
 \cW=S\tilde{p}^2 + \mu_0 \mathfrak{m}_{2N-1}^2 + \tilde{p}\tilde{q} \mathfrak{m}_{2N-1}+ \\
 \mathfrak{m}_0^2\,\beta (\text{det}(S)\mu_0 + S^{2N-3}\tilde{q}\tilde{q})
  \ea \ee
where we expanded the cubic and determinant terms in the r.h.s. of \eqref{eq:SconfW=0} as:
\be
q (\mathbb{S}) q\,=\,
\left(\begin{matrix} \tilde{p} \,&\,0\,& \, \mathfrak{m}_{2N-1} \end{matrix}\right)         
     \left(\begin{matrix}S \,&\,0\,&\,\tilde{q}\, \\
                 0   \, & \beta& 0\\
                 \tilde{q} \, & 0 & \mu_0
                 \end{matrix}\right)
\left(\begin{matrix} \tilde{p} \\0\\ \mathfrak{m}_{2N-1}  \end{matrix}\right)                          
                 =\, S\tilde{p}^2 + \mu_0 \mathfrak{m}_{2N-1}^2 + \tilde{p}\tilde{q} \mathfrak{m}_{2N-1}  \ee

\be
\text{det}(\mathbb{S}) \,=\,
\text{det} \left(\begin{matrix}S \,&\,0\,&\,\tilde{q}\, \\
                 0   \, & \beta& 0\\
                 \tilde{q} \, & 0 & \mu_0
                 \end{matrix}\right)\,=\,
                \beta \epsilon_{2N-2} \epsilon_{2N-2} (S^{2N-2} \mu_0 +  S^{2N-3}\tilde{q}\tilde{q}) 
\ee

Applying duality \eqref{eq:SconfW=01} to $\cT_{2'}$, four gauge singlets appear:
\begin{itemize}
\item A singlet $\beta$, dual of $\Tr(uu)$. Observe that $R[\beta]=2Nr_\Phi$. $\beta$ maps to $\text{Tr}(\Phi^{2N})$ in the original theory $\cT_1$.
\item A singlet $\mu_0$, dual of $\Tr(tt)$. It has R-charge $R[\mu_0]=2r_q+r_\Phi$ and it maps to $\text{Tr}(\Phi q q)$ in $\cT_1$.
\item A singlet $\mathfrak{m}_{2N-1}$ dual of the baryon monopole $\mathfrak{M}^{o,-}_{\epsilon c^{2N-2}u}$. It maps to the dressed monopole $\mathfrak{M}_{\Phi^{2N-1}}$ in $\cT_1$.
\item A singlet $\mathfrak{m}_0$ dual of the monopoles $\mathfrak{M}^{0,+}$. It maps to the fundamental monopole $\M$ in $\cT_1$.
\end{itemize}
The linear monopole term in the superpotential is lifted, and we are left with
\be 
 \scalebox{0.9}{
\bpic  
\path (-5,0) node{$\CT_3:$} -- (-3,0) node[blue](x1) {$C_{N-1}$} -- (-3,-1.3) node[rectangle,draw](x2) {$1$}-- (-3+1.3,0) node[rectangle,draw](x3){$1$} -- 
(1,-0.5) node{+ 4 singlets} -- (1, -1) node{$\{\mu_0, \mathfrak{m}_0, \mathfrak{m}_{2N-1}, \beta\}$};
\draw[-] (x1) edge [out=55,in=125,loop,looseness=3] node[above]  {$S$}(x1);
\node at (-2,-2.3){$ \cW=\Tr(\tilde{p}\,S\tilde{p})+\mu_0 \mathfrak{m}_{2N-1}^2 +  \text{Tr}(\tilde{p}\tilde{q}) \mathfrak{m}_{2N-1} +
 \mathfrak{m}_0^2\,\beta (\text{det}(S)\mu_0 + \epsilon \epsilon S^{2N-3}\tilde{q}\tilde{q})$};
\draw[-] (x1) edge node[right]{$\tilde q$} (x2);
\draw[-] (x1) edge node[above]{$\tilde p$} (x3);
 \epic }
 \ee
Notice that $\cT_3$ is equal to $\cT_1$, modulo replacing $N \rightarrow N-1$ and modulo superpotential terms which are of \emph{flipping type}. This means that we can perform the same procedure that took us from $\mathcal{T}_1$ to $\mathcal{T}_3$ $N-1$ times, and one is left with a theory with trivial gauge group and $4N$ singlets, that is a Wess-Zumino model, as promised.

\paragraph{S-confinement of the flipped $Sp(N)$ with adjoint and $2$ fundamentals}
We can repeat the above procedure starting from $\cT_1$ with the $N$ traces of the adjoint $\text{Tr}(\Phi^{2i})$ flipped by $\rho_i$.
\be\label{T1f}
 \scalebox{0.9}{
\bpic  
\path (-5,0) node{$\CT_{1\, flipped}:$} -- (-3,0) node[blue](x1) {$C_N$} -- (-3,-1.3) node[rectangle,draw](x2) {$1$}-- (-3+1.3,0) node[rectangle,draw](x3){$1$};
\draw[-] (x1) edge [out=55,in=125,loop,looseness=3] node[above]  {$\Phi$}(x1);
\node at (3 , 0){$ \cW=\Tr(p\Phi p)+\sum_{J=1}^{N}\rho_J\Tr(\Phi^{2J})$};
\draw[-] (x1) edge node[right]{$q$} (x2);
\draw[-] (x1) edge node[above]{$p$} (x3);
 \epic }
 \ee
 The superpotential term $\sum_{J=1}^{N}\rho_J\Tr(\Phi^{2J})$ maps to $\sum_{J=1}^{N}\rho_J\Tr(A^{2J})$ in $\cT_{2, flipped}$, to $\sum_{J=1}^{N}\rho_J \Tr((cc)^{2J})$ in $\cT_{2', flipped}$ and to $\rho_N \beta + \sum_{J=1}^{N-1}\rho_J\Tr(S^{2J})$ in $\cT_{3, flipped}$. So when we arrive to $\cT_3$, and produce the singlet $\beta$, $\beta$ pairs up with $\rho_N$, the flipper of $\text{Tr}(\Phi^{2N})$: both singlets become massive and disappear. As a consequence of this, the superpotential in $\cT_3$ simplifies to:
\be 
 \scalebox{0.9}{
\bpic  
\path (-5,0) node{$\CT_{3,\,flipped}:$} -- (-3,0) node[blue](x1) {$C_{N-1}$} -- (-3,-1.3) node[rectangle,draw](x2) {$1$}-- (-3+1.3,0) node[rectangle,draw](x3){$1$} -- 
(1,-0.5) node{+ 3 singlets} -- (1, -1) node{$\{\mu_0, \mathfrak{m}_0, \mathfrak{m}_{2N-1}\}$};
\draw[-] (x1) edge [out=55,in=125,loop,looseness=3] node[above]  {$S$}(x1);
\node at (-2,-2.3){$ \cW=\Tr(\tilde{p}\,S\tilde{p})+\mu_0 \mathfrak{m}_{2N-1}^2+\sum_{J=1}^{N-1}\rho_J\Tr(\Phi^{2J})+  \text{Tr}(\tilde{p}\tilde{q}) \mathfrak{m}_{2N-1} $};
\draw[-] (x1) edge node[right]{$\tilde q$} (x2);
\draw[-] (x1) edge node[above]{$\tilde p$} (x3);
 \epic }
 \ee
 Iterating the procedure $N$ times we get the duality with the cubic Wess-Zumino\footnote{To be precise, this argument is not enough to prove that the final superpotential is exactly the one on the r.h.s. of \eqref{eq:SconfSPflip}. See \cite{BB1} for a full proof of the form of the cubic Wess-Zumino superpotential in the case of $4d$ $\cN=1$ $Usp(2N)$ with antisymmetric and $6$ fundamentals.}
 \be
\label{eq:SconfSPflip}
\ba{c} Sp(N) \, \text{w/ adjoint $\Phi$}\\
   \text{and $2$ fundamentals\,$q$, $p$} \\ 
   \cW= \Phi p p +  \sum_{J=1}^{N}\rho_J\Tr(\Phi^{2J}) \ea
     \qquad      \Longleftrightarrow \qquad
\ba{c} \text{Wess-Zumino w/\,}  3N \,\text{chirals} \\
 \mathfrak{m}_j \leftrightarrow \M_{\Phi^j},\, j=0,\ldots,2N-1 \\
 \mu_i \leftrightarrow \text{Tr}(\Phi^{2i+1} qq),\, i=0,\dots,N-1\,,  \\
    \mathcal{W} = \sum_{i,j,k} \mathfrak{m}_i \mathfrak{m}_j \mu_k \delta_{i+j+2k-4N+2}
  \ea \ee
  Using \eqref{chargesSconf}, we see that the superpotential on the r.h.s. is the most general consistent with the $U(1)_\Phi \times U(1)_q \times U(1)_R$ global symmetry.

Notice that we didn't include the mapping of the operator $\text{Tr}(p q)$. The reason is that $\text{Tr}(p q)$ is actually a composite operator: $\text{Tr}(p q) \sim \sum_{i=1}^N \text{Tr}(q \Phi^{2i-1} q) \times \M_{\Phi^{2N-2i}}$.\footnote{Notice that this chiral ring relation is consistent with the global symmetries, as can be seen from \eqref{chargesSconf}: the l.h.s. has $R=1-\frac{r_\Phi}{2}+r_q$, and the r.h.s. has $R=2r_q+(2i-1) r_\Phi +1-r_q-(2N-1/2)r_\Phi+(2N-2i)r_\Phi$, which are equal.}

\subsection{$SO(2N)$ with adjoint and $1$ fundamental}
Starting from 
\be 
 \scalebox{0.9}{
\bpic  
\path (-5,0) node{$\CT_1:$} -- (-3,0) node[red](x1) {$D_N$} -- (-3,-1.3) node[rectangle,draw](x2) {$1$};
\draw[-] (x1) edge [out=55,in=125,loop,looseness=3] node[above]  {$A$}(x1);
\node at (0.5,-0.6){$ \cW=0$};
\draw[-] (x1) edge node[right]{$t$} (x2);
 \epic }
 \ee
we deconfine the adjoint to:
\be
\scalebox{1}{
\begin{tikzpicture}[baseline]
\tikzstyle{every node}=[font=\footnotesize]
\node[draw=none, blue] (node1) at (0,0) {$C_{N-1}$};
\node[draw=none, red] (node2) at (2,0) {$D_N$};
\node[draw, rectangle] (sqnode) at (2,-1.2) {$1$};
\draw[black,solid] (node1) edge node[below]{$c$} (node2);
\draw[black,solid] (node2) edge node[right]{$t$} (sqnode);
\node at (5,-0.2){$ \cW=\gamma \mathfrak{M}^{\bullet,o}$};
\node[draw=none] at (-2,0) {$\mathcal T_{1'}:$};
\end{tikzpicture}}
\ee
The map of the operators is not straightforward in this case but there is a non-trivial mixing. Let us make an example: the flipping field $\gamma$ maps to the baryon $\text{Pf}A$ in $\cT_1$ and the monopole $\cM^{0,+}$ in $\cT_2$ is dual to the baryon monopole in the original theory.
We use duality \eqref{eq:SconfW=0} to confine the $SO(2N)$ node and get
\be 
 \scalebox{0.9}{
\bpic  
\path (-5,0) node{$\CT_{2}:$} -- (-3,0) node[blue](x1) {$C_{N-1}$} -- (-3,-1.3) node[rectangle,draw](x2) {$1$}-- (-3+1.3,0) node[rectangle,draw](x3){$1$} -- 
(1,-0.5) node{+ 4 singlets} -- (1, -1) node{$\{\gamma, \mu_0, \mathfrak{m}_{2N-2}, \mathfrak{m}_{-}\}$};
\draw[-] (x1) edge [out=55,in=125,loop,looseness=3] node[above]  {$S$}(x1);
\node at (-2,-2.3){$ \cW=\Tr(\tilde{p}\,S\tilde{p})+\mu_0 \mathfrak{m}_{2N-2}^2+ \text{Tr}(\tilde{p}\tilde{q}) \mathfrak{m}_{2N-2} +$};
\node at (-2,-2.9){$(\mathfrak{m}_{-})^2 (\mu_0 \text{det}(S) + \epsilon \epsilon S^{N-2}\tilde{q}\tilde{q})+\gamma (\dots)$};
\draw[-] (x1) edge node[right]{$\tilde q$} (x2);
\draw[-] (x1) edge node[above]{$\tilde p$} (x3);
 \epic }
 \ee
The four singlets $\{\gamma, \mu_0, \mathfrak{m}_{2N-2}, \mathfrak{m}_{-}\}$ map to $\{\epsilon A^N, \text{Tr}(tt), \M^+_{A^{2N-2}}, \M^-_{\epsilon A^{N-1}}\}$ in $\cT_1$. The flipping field $\gamma$ appears in the superpotential multiplying all possible cubic polynomials involving two dressed monopoles and a meson such that the sum of all R-charges adds up to 2. Since the precise form of such superpotential is not useful in the following, we do not write it explicitly.
Using the result of section \ref{sec:SconfSP}, we can replace the $Sp(N-1)$ with a Wess-Zumino model with $4(N-1)$ fields, and obtain a Wess-Zumino model with $4+4(N-1)=4N$ fields ($N-1$ powers of the adjoint, $2N-1$ dressed monopoles, $N$ dressed mesons, a baryon and a baryon-monopole). 

Flipping the traces of the powers of the adjoint, the baryon-monopole and the baryon, we get a unitary S-confining duality with a Wess-Zumino model with a cubic superpotential:
\be
\label{eq:SconfSOeflip}
\ba{c} SO(2N)\,\,\text{with adjoint}\,\,A \\
 \text{and $1$ flavor}\,\,t,\\
\cW=\sum_{J=1}^{N-1}\rho_J\Tr(A^{2J})+\\
\gamma \, \mathfrak{M}^{-}_{\epsilon_{2N-2}A^{N-1}} + \sigma\, \epsilon_{2N} A^{N}   \ea
     \qquad      \Longleftrightarrow \qquad
\ba{c} \text{Wess-Zumino with $3N-1$ chiral fields}\\
 \mathfrak{m}_j \leftrightarrow \M^+_{A^j},\, j=0,\ldots,2N-2 \\
 \mu_i \leftrightarrow \text{Tr}(t A^{2i} t),\, i=0,\dots,N-1\,,  \\
     \mathcal{W} = \sum_{i,j,k} \mathfrak{m}_i \mathfrak{m}_j \mu_k \delta_{i+j+2k-4N+4}\,.  \ea \ee
It is easy to check that the superpotential on the r.h.s. preserves the $U(1)_t \times U(1)_A \times U(1)_R$ global symmetry using the formula for the R-charges of the monopole in $\cT_1$:
\be R[\M^+] = 1 - r_t  - (2N-2) r_A \,. \ee

\subsection{$SO(2N+1)$ with adjoint and $1$ fundamental}
For completeness, we discuss this case explicitly, but all the ingredients already appeared in the discussion for $Sp(N)$ with adjoint and $2$ fundamentals of section \ref{sec:SconfSP}.

Starting from
\be 
 \scalebox{0.9}{
\bpic  
\path (-5,0) node{$\CT_1:$} -- (-3,0) node[red](x1) {$B_N$} -- (-3,-1.3) node[rectangle,draw](x2) {$1$};
\draw[-] (x1) edge [out=55,in=125,loop,looseness=3] node[above]  {$A$}(x1);
\node at (0.5,-0.6){$ \cW=0$};
\draw[-] (x1) edge node[right]{$t$} (x2);
 \epic }
 \ee
we deconfine the adjoint to:
\be
\scalebox{1}{
\begin{tikzpicture}[baseline]
\tikzstyle{every node}=[font=\footnotesize]
\node[draw=none, blue] (node1) at (0,0) {$C_{N-1}$};
\node[draw=none, red] (node2) at (2,0) {$B_N$};
\node[draw, rectangle] (sqnode) at (2,-1.2) {$1$};
\node[draw, rectangle] (auxnode) at (1,1.2) {$1$};
\draw[black,solid] (node1) edge node[below]{$c$} (node2);
\draw[black,solid] (node2) edge node[right]{$t$} (sqnode);
\draw[black,solid] (node2) edge node[right]{$u$} (auxnode);
\draw[black,solid] (node1) edge node[left]{$v$} (auxnode);
\node at (5,-0.2){$ \cW=\mathfrak{M}^{\bullet,o}+\Tr(vcu)$};
\node[draw=none] at (-2,0) {$\mathcal T_{1'}:$};
\end{tikzpicture}}
\ee
We use duality \eqref{eq:SconfW=0} to confine the $SO(2N+1)$ node, and lift the linear monopole superpotential. Modulo flips, we obtain $Sp(N-1)$ with adjoint $\Phi$ and $2$ fundamentals $p,q$, $\cW= \Phi p p $. The flippers are $6$ singlets mapping to 
$$\{\text{Tr}(uu), \text{Tr}(tt), \text{Tr}(ut),  \M^{0,-}_{\epsilon \cdot c^{2N-2}u},  \M^{0,-}_{\epsilon \cdot c^{2N-2}t}, \M^{0,+}\}$$
 in $\mathcal T_{1'}$. These $6$ singlets map to 
 $$\{\text{Tr}(A^{2N}),\text{Tr}(tt), \epsilon A^N t,  \M^{+}_{A^{2N-1}},  \M^{-}_{\epsilon \cdot A^{N-1}t}, \M^{+}\}$$
  in $\mathcal T_{1}$. Using the result of section \ref{sec:SconfSP}, we can replace the $Sp(N-1)$ with a Wess-Zumino model with $4(N-1)$ fields, and we obtain a Wess-Zumino model with $6+4(N-1)=4N+2$ fields ($N$ powers of the adjoint, $2N$ dressed monopoles, $N$ dressed mesons, a baryon and a baryon-monopole).

Flipping the traces of the powers of the adjoint, the baryon-monopole and the baryon, we get a unitary S-confining duality with the same Wess-Zumino model dual to $Sp(N)$ with adjoint and $2$ fundamentals, with a cubic superpotential:
\be
\label{eq:SconfSOoflip}
\ba{c} SO(2N+1)\,\,\text{with adjoint}\,\,A \\
 \text{and $1$ flavor}\,\,t,\\
\cW=\sum_{J=1}^{N}\rho_J\Tr(A^{2J})+\\
\gamma \, \mathfrak{M}^{-}_{\epsilon_{2N-1}A^{N-1}t} + \sigma\, \epsilon_{2N+1} A^{N} t  \ea
     \qquad      \Longleftrightarrow \qquad
\ba{c} \text{Wess-Zumino with $3N$ chiral fields}\\
 \mathfrak{m}_j \leftrightarrow \M^+_{A^j},\, j=0,\ldots,2N-1 \\
 \mu_i \leftrightarrow \text{Tr}(t A^{2i} t),\, i=0,\dots,N-1\,,  \\
     \mathcal{W} = \sum_{i,j,k} \mathfrak{m}_i \mathfrak{m}_j \mu_k \delta_{i+j+2k-4N+2}\,.  \ea \ee
It is easy to check that the superpotential on the r.h.s. preserves the $U(1)_t \times U(1)_A \times U(1)_R$ global symmetry using the formula for the R-charges of the monopole in $\cT_1$:
\be R[\M^+] = 1 - r_t  - (2N-1) r_A \,. \ee

\section{$3d$ mirror of $A_{2N}$ Argyres-Douglas from $4d$ Lagrangians} \label{sec:3dmirrors}
$4d$ $\cN=1$ Lagrangians flowing in the IR to  $\cN=2$ $(A_1, A_{2N})$ Argyres-Douglas theories have been discovered in \cite{Maruyoshi:2016aim} (see \cite{Agarwal:2016pjo, Agarwal:2017roi, Benvenuti:2017bpg} for generalizations). In four dimensions, the $\cN=2$ Argyres-Douglas theory is reached in the IR, at the end of an RG flow described by an $\cN=1$ lagrangian:
 \be
\label{eq:ArgyresDouglas4D}
\ba{c} \scalebox{0.9}{$4d\,\,\,Sp(N)\,\,\text{with adjoint}\,\,\Phi$}\\
\scalebox{0.9}{$\text{and 2 flavors}\,\,p,\,q,$}\\
\scalebox{0.9}{$\cW=\Tr(p\Phi p)+\sum_{J=1}^{N}\rho_J\Tr(\Phi^{2J})+$}\\
\scalebox{0.9}{$+\sum_{J=1}^N\tau_J\Tr(q\Phi^{2J-1}q)$} \ea
     \qquad      \Longrightarrow \qquad
\ba{c}\scalebox{1}{$(A_1, A_{2N})$\,\, Argyres-Douglas}
  \ea \ee

\cite{Benvenuti:2018bav} argued that, reducing the theory to $3d$, the superpotential of the lagrangian theory does not change. This behaviour is similar to the case of $(A_1, A_{2N+1})$ Argyres-Douglas, discussed in \cite{Benvenuti:2017kud}, but notice that for $(A_1, D_{2N})$ and $(A_1, D_{2N+1})$ the $3d$ superpotential is different from the $4d$ superpotential \cite{Benvenuti:2017bpg, Agarwal:2018oxb}.

\cite{Benvenuti:2018bav} also proposed that the $3d$ mirror of $(A_1, A_{2N})$ Argyres-Douglas is a theory of $N$ free massless hypermultiplets. Such a statement implies  the duality between $4$-supercharges lagrangian and a theory of free chirals:
 \be
\label{eq:ArgyresDouglasSp}
\ba{c} \scalebox{0.9}{$Sp(N)\,\,\text{with adjoint}\,\,\Phi$}\\
\scalebox{0.9}{$\text{and 2 flavors}\,\,p,\,q,$}\\
\scalebox{0.9}{$\cW=\Tr(p\Phi p)+\sum_{J=1}^{N}\rho_J\Tr(\Phi^{2J})+$}\\
\scalebox{0.9}{$+\sum_{J=1}^N\tau_J\Tr(q\Phi^{2J-1}q)$} \ea
     \qquad      \Longleftrightarrow \qquad
\ba{c}\scalebox{0.9}{$ 2N\,\,\text{free chiral fields}$}\\
\scalebox{0.9}{$\sim\,Sp(N)\text{-monopoles}$}
  \ea \ee
\cite{Benvenuti:2018bav} checked this duality with the $S^3$ partition function. See \cite{Dedushenko:2019mnd} for additional checks of the $3d$ mirror duality. With the techniques developed in this paper, we are now in a position to \emph{derive} duality \eqref{eq:ArgyresDouglasSp}. Duality \eqref{eq:ArgyresDouglasSp} is an immediate consequence of duality \eqref{eq:SconfSPflip}. From \eqref{eq:SconfSPflip}:
 \be
\ba{c} Sp(N) \, \text{w/ adjoint $\Phi$}\\
   \text{and $2$ fundamentals\,$q$, $p$} \\ 
   \cW= \Phi p p +  \sum_{J=1}^{N}\rho_J\Tr(\Phi^{2J}) \ea
     \qquad      \Longleftrightarrow \qquad
\ba{c} \text{Wess-Zumino w/\,}  3N \,\text{chirals} \\
 \mathfrak{m}_j \leftrightarrow \M_{\Phi^j},\, j=0,\ldots,2N-1 \\
 \mu_i \leftrightarrow \text{Tr}(\Phi^{2i+1} qq),\, i=0,\dots,N-1\,,  \\
    \mathcal{W} = \sum_{i,j,k} \mathfrak{m}_i \mathfrak{m}_j \mu_k \delta_{i+j+2k-4N+2}
  \ea \ee
we flip the l.h.s. mesons and their r.h.s. images
\be   \text{Tr}(\Phi^{2i+1} qq) \leftrightarrow \mu_i \,.\ee
On the r.h.s we are left with a theory of $2N$ chiral fields  $\mathfrak{m}_j \leftrightarrow \M_{\Phi^j}$, $j=0,\ldots,2N-1$. Such a theory is free because no non-trivial polynomial superpotential invariant under $U(1)_\Phi \times U(1)_{q} \times U(1)_R$ exists. Hence, we obtain the $3d$ mirror duality \eqref{eq:ArgyresDouglasSp} from the S-confining duality \eqref{eq:SconfSPflip}. Notice that in \eqref{eq:ArgyresDouglasSp} the global $SU(4N)$ symmetry rotating the $4N$ chiral fields (or $2N$ hypers) is emergent in the IR, it is not visible in the UV, not even the Cartan generators are visible. This phenomenon is the same as for the theory $SU(N)$ with adjoint and $1$ flavor with appropriate flipping superpotential, which is dual to $\mathcal{N}=4$ U(1) with $N$ flavors \cite{Benvenuti:2017kud} and hence has $SU(N)$ global symmetry in the IR, which is not visible in the UV and is destroyed if we  start in the UV from a different flipping type superpotential.

\section{Generalized duality appetizers for $SO$ or $Sp$ with adjoint} \label{sec:appetizers}
Some time ago, \cite{Jafferis:2011ns} noticed that $\cN=2$ $SU(2)_1$ with an adjoint and vanishing superpotential is dual to a theory of a free chiral field, and called this duality \emph{duality appetizer}. \cite{Kapustin:2011vz} soon generalized the duality to a class of dualities, relating an $\cN=2$ theory with a simple gauge group, appropriate Chern-Simons level, a rank-$2$ field and vanishing superpotential to theories of free chiral fields. In the case of $SO(2N)_1$  with adjoint, we can prove such duality appetizers as a simple consequence of the S-confining dualities obtained in section \ref{sec:Sconf}.\footnote{The duality appetizers for $U(N)$ with adjoint and for $Sp(N)$ with antisymmetric can be obtained in a similar way from S-confining dualities, see \cite{Amariti:2014lla, Benvenuti:2018bav}.} A similar strategy allows us to prove  an appetizer duality for $SO(2N+1)_1$ with adjoint, and the duality appetizer proposed in \cite{Benvenuti:2018bav} for $Sp(N)_{\frac{1}{2}}$ with adjoint and fundamental.

\paragraph{$Sp(N)_{\frac{1}{2}}$ with adjoint and fundamental}
We start from \eqref{eq:Sconf0} and turn on a real mass for the flavor $q$. The operators $\M_{\Phi^j}$ and $\text{Tr}(tq\Phi^{2i+1} q)$  are charged under the corresponding $U(1)$ symmetry, the operators $\text{Tr}(\Phi^{2k})$ are uncharged. 

On the l.h.s., in the IR we are left with $Sp(N)_{\frac{1}{2}}$ with adjoint and fundamental $p$. On the r.h.s., in the IR, we are left with a theory with the singlets $\sigma_k \leftrightarrow \text{Tr}(\Phi^{2k})$. No non-trivial polynomial superpotential invariant under $U(1)_{\Phi} \times U(1)_R$ exists, so the dual must be free:
\be
\ba{c} Sp(N)_{\frac{1}{2}} \, \text{w/ adjoint $\Phi$}\\
 \text{and fundamental} \,\, p\\
   \cW= \text{Tr}(p \Phi p) \ea
     \qquad      \Longleftrightarrow \qquad
\ba{c} N\,\, \text{free chirals} \\
 \sigma_k \leftrightarrow \text{Tr}(\Phi^{2k}),\, k=1,\ldots,N.
  \ea \ee
This duality was conjectured in \cite{Benvenuti:2018bav}.

\paragraph{$SO(2N)_1$ with adjoint}
We start from \eqref{eq:Sconf2} and turn on a real mass for the flavor $t$. The operators $\M_{\Phi^j}$, $\text{Tr}(t \Phi^{2i} t)$ and $\M^-_{\Phi^{N-1}}$ are charged under the corresponding $U(1)$ symmetry, the operators $\text{Tr}(\Phi^{2k})$ and $\text{Pf}(\Phi)$ are uncharged.

On the l.h.s., in the IR we are left with $SO(2N)_1$ with an adjoint $\Phi$ and $\cW=0$. On the r.h.s., in the IR, we are left with a theory with the singlets $\sigma_k \leftrightarrow \text{Tr}(\Phi^{2k})$ and $\mathbb{B} \leftrightarrow Pf(\Phi)$. No non-trivial polynomial superpotential invariant under $U(1)_{\Phi} \times U(1)_R$ exists, so the dual must be free:
\be
\ba{c} SO(2N)_1 \, \text{w/ adjoint $\Phi$}\\
   \cW= 0 \ea
     \qquad      \Longleftrightarrow \qquad
\ba{c} N\,\, \text{free chirals} \\
 \sigma_k \leftrightarrow \text{Tr}(\Phi^{2k}),\, k=1,\ldots,N-1\\
 \mathbb{B} \leftrightarrow \text{Baryon}\,\,\text{Pf}\,\Phi\,.
  \ea \ee
This duality was conjectured in \cite{Kapustin:2011vz}.

\paragraph{$SO(2N+1)_1$ with adjoint}
We start from \eqref{eq:Sconf3} and turn on a real mass for the flavor $t$. The operators $\M_{\Phi^j}$, $\text{Tr}(t \Phi^{2i} t)$, $\M^-_{t\Phi^{N-1}}$ and the baryon $\epsilon \cdot t \Phi^N$ are charged under the corresponding $U(1)$ symmetry, the operators $\text{Tr}(\Phi^{2k})$ are uncharged. At the end of the RG flow, the baryon-monopole $\M^-$ transforms non-trivially under charge conjugation and transforms non-trivially as a vector under the $SO(2N+1)$ gauge group, because of the non-vanishing Chern-Simons level. The dressed monopole $\M^-_{\Phi^{N}}$ is however still gauge invariant and belongs to the chiral ring. 
A way to check the presence of this particular gauge-invariant baryon monopole is through the use of the supersymmetric index:
\begin{equation}
\label{eq:IndexEl3}
\mathcal{I}_{SO(2N+1)_1\,\text{w/}\,\Phi}\,=\,1+\phi^2\,x^{2R_\Phi}+\omega\,\phi^{-1}\,x^{1-NR_\Phi}+\dots\,,
\end{equation}
where $\phi$ is the fugacity for the $U(1)$ global symmetry rotating the adjoint field and $\omega$ is the fugacity for the magnetic symmetry. The first non-trivial term in \eqref{eq:IndexEl3} corresponds to the operator $\text{Tr}\Phi^2$ while the second corresponds to a monopole, given the non-trivial charge under $\mathcal{M}$. The R-charge is consistent with the hypothesis that the operator appearing in the index is $\mathfrak{M}^-_{\Phi^N}$ as claimed.

 We flip this particular operator adding an extra flipping field ${\rho}$.

On the l.h.s., in the IR we are left with $SO(2N+1)_1$ with an adjoint $\Phi$ and $\cW=0$. On the r.h.s., in the IR, we are left with a theory with the singlets $\sigma_k \leftrightarrow \text{Tr}(\Phi^{2k})$. No non-trivial polynomial superpotential invariant under $U(1)_{\Phi} \times U(1)_R$ exists, so the dual must be free:
\be
\label{eq:appetizerOdd}
\ba{c} SO(2N+1)_1 \, \text{w/ adjoint $\Phi$}\\
   \cW= \rho\,\M^-_{\Phi^{N}}  \ea
     \qquad      \Longleftrightarrow \qquad
\ba{c} N\,\, \text{free chirals} \\
 \sigma_k \leftrightarrow \text{Tr}(\Phi^{2k}),\, k=1,\ldots,N\,.
  \ea \ee
  Because of the non-trivial superpotential on l.h.s, in the case of $SO(2N+1)_1$ one does not obtain a proper appetizer.

\acknowledgments
We are grateful to Ivan Garozzo for collaboration in the initial stage of this work and invaluable discussions. We thank Francesco Benini for useful conversations. We are grateful to Matteo Sacchi for useful remarks about the duality \eqref{eq:appetizerOdd}. SB is partially supported by the INFN Research Project GAST. GLM is supported by the ERC Consolidator Grant number 772408 ``String landscape''.


\appendix

\section{$3d$ Supersymmetric index} \label{app:index}

All the dualities proposed in the main text have been checked computing the supersymmetric index, which is an RG-invariant quantity that can be computed for different 3d $\mathcal{N}=2$ quiver theories and that must match across a duality. Its formal definition involves a trace over the Hilbert space of the theory on 
$S^2 \times \mathbb R$ \cite{Bhattacharya:2008zy, Bhattacharya:2008bja, Kim:2009wb, Imamura:2011su, Kapustin:2011jm, Dimofte:2011py}, (we use the definitions of \cite{Aharony:2013dha, Aharony:2013kma}):
\be\label{TrInd}
	\mathcal I(x, \vec\mu)=\Tr \left[ (-1)^{J_3} x^{\Delta + J_3} \prod_{i} \mu_i^{q_i} \right],
\ee
where the various quantities in the formula represents
\begin{itemize}
	\item{$\Delta$: is the energy whose scale is set by the radius of $S^2$,}
	\item{$J_3$: is the Cartan generator for the $SO(3)$ isometry of the $S^2$,}
	\item{$\mu_i, \, q_i$: respectively the fugacities and charges of the global non-$R$ symmetries.}
	\item{$R$ will denote the R-charge in the following.}
\end{itemize}
The only non trivial contributions to the index comes from states that are annihilated by two supercharges and satisfy the relation
\be
	\Delta=R+J_3.
\ee
It is not so easy to employ the definition \eqref{TrInd} to perform an explicit computation of the index; here the localization techniques come at rescue. Indeed, the index can be computed as the partition function on $S^2 \times S^1$ given by the following expression
\be\label{LocInd}
	\mathcal I(x)=\sum_{\bold m} \frac{1}{|\mathcal W_{\bold m}|} \int
	\frac{d \bold z}{2 \pi i \bold z} Z_{\text{cl}} \, Z_{\text{vec}} \, Z_{\text{mat}},
\ee
where the integral is taken over the Cartan torus of the gauge group whose fugacities are $\bold z$, $|\mathcal W_{\bold m}|$ is the dimension of the Weyl group that is left unbroken by the monopole background specified by the GNO magnetic fluxes $\bold m$. Localization implies that the only non trivial contribution to \eqref{LocInd} from non-exact term in the classical action and  
from $1$-loop terms. The various terms $Z_{\text{cl}}, \, Z_{\text{vec}}, \, Z_{\text{mat}}$ have the following expressions
\begin{itemize}
	\item{$Z_{\text{cl}}$: The classical terms includes only CS couplings and, more generally, BF 	terms. Take a gauge group whose rank is $\text{rk}G$. Denoting the fugacity for the 		topological with $\omega$ and the associated flux as $\bold n$, and given a level $k$ CS term we have
	\be
		Z_\text{cl}=\prod_{i=1}^{\text{rk}G} \omega^{m_i} z_i^{k\, m_i+\bold n}
	\ee
	The topological symmetry is only present for $U(N)$ gauge groups and for $SO(N)$ gauge groups. In the first former case, the topological symmetry is $U(1)$, while in the latter case the topological symmetry is a discrete $\mathbb{Z}_2$ group (implying that the condition $\omega^2=1$ must be enforced).\footnote{In the case of $SO(2)$, the topological symmetry is still $U(1)$, being $SO(2)=U(1)$.}}
	\item{$Z_{\text{vec}}$: The contribution for an $\mathcal N=2$ vector multiplet reads
	\be
		Z_{\text{vec}}^{G}(\mathbf{z})=\prod_{\alpha \in \mathfrak g} x^{-\frac{|\alpha(m)|}{2}} (1-(-1)^{\alpha(m)}
		\bold z^\alpha x^{|\alpha(m)|})\,,
	\ee
	where we denoted by $\alpha$ the weights of the adjoint representation of the gauge group G.}
	\item{$Z_{\text{mat}}$: The contribution of an $\mathcal N=2$ chiral multiplet with $R$-charge $r$ transforming in the representations $\mathcal R$ and $\mathcal R_F$ under the gauge and flavour group, whose weights we denote as $\rho, \rho_F$, is
	\begin{align}
		Z_{\text{chi}}(\mathbf{z}_{\mathcal{R}},\boldsymbol{\mu}_{\mathcal{R}_F})=\prod_{\rho \in \mathcal R} \prod_{\rho_F \in \mathcal R_F} 
		&(z^\rho \mu^{\rho_F} x^{r-1})^{-\frac{|\rho(m)+\rho_F(n)|}{2}} \times \\
		&\times \frac{((-1)^{\rho(m)+\rho_F(n)} z^{-\rho} \mu^{-\rho_F} x^{2-r+|\rho(m)+\rho_F(n)|}; x^2)_\infty}{((-1)^{\rho(m)+\rho_F(n)} z^{-\rho} \mu^{-\rho_F} x^{r+|\rho(m)+\rho_F(n)|}; x^2)_\infty}.
	\end{align}}
\end{itemize}

Explicit expressions for $Z_{\text{chi}}$ and $Z_{\text{vec}}$ for both ortho-symplectic groups and unitary ones can be found in the appendices of \cite{Beratto:2021xmn,Mekareeya:2022spm}.

\section{Monopoles and dualities for orthogonal gauge groups}
\label{sec:reviewSO}
In this section we will review the current knowledge about the monopole operators in theories involving orthogonal gauge groups and the related Seiberg-like dualities proposed in \cite{Aharony:2011ci,Aharony:2013kma}. 

Let us start considering an $SO(N)$ theory with $F$ flavors $Q$. The global symmetry group of this model is:
\be
\label{eq:SOglobalsym}
G_{N,F}\,=\,(U(F)\times \mathbb{Z}^\cC_2\times \mathbb{Z}_2^\cM)/\mathbb{Z}_2\,,
\ee
where the Abelian factor in $U(F)$ is the $U(1)_Q$ axial symmetry acting on the chiral fields; the discrete $\mathbb{Z}_2^\cC$ factor is the charge conjugation symmetry, whose non-trivial element consists of the orthogonal transformation (in $O(N)$) with determinant equal to $-1$, {\it i.e.} a reflection; the magnetic discrete symmetry $\mathbb{Z}_2^\cM$, instead, acts on the Coulomb branch coordinates charging $-1$ the fundamental monopole operators.\footnote{The two discrete factors $\mathbb{Z}_2^{\cC,\cM}$ and the element $e^{i\pi}$ of $U(1)_Q$ are not really independent but they actually satisfy the relation $e^{i\pi}_Q\cdot \cC^N\cdot \cM^F=1$ \cite{Aharony:2013kma}; this is the reason why a common $\mathbb{Z}_2$ factor is mod out in \eqref{eq:SOglobalsym}.} As usual, on the Coulomb branch the gauge group is broken to the Cartan $U(1)^{r_N}$ with $r_N=\lfloor N/2\rfloor$. Semi-classically, the basic monopole operators can be written as:
\be
\cV_{\pm}\,\approx\,e^{\pm\left(\tfrac{\alpha_1}{g^2}+i\phi_1\right)}\,,
\ee
where we denoted with $\alpha_i$ and $\phi_i$ the dual photon and adjoint scalar respectively for the $i^{th}$ Abelian vector multiplet in $U(1)^{r_N}$. Charge conjugation acts non-trivially on the two monopoles $\cV_{\pm}$ swapping them, so that it is useful to define the even and odd $\mathbb{Z}_2^\cC$ combinations
\be
\mathfrak M^{\pm}\,=\,\cV_{+}\,\pm\, \cV_-\,.
\ee
Observe that both the monopoles breaks the gauge group down to $S\left(O(N-2)\times O(2)\right)$, including the transformation with $-1$ determinant in both the $O(N-2)$ and the $O(2)$ factors. In particular, in order for the monopole to be gauge invariant, it must be invariant under charge conjugation in the $O(2)$ factor; following the previous discussion, only $\mathfrak{M}^+$ has this property, while $\mathfrak M^-$ is not gauge invariant on its own. 

However, we can still build a gauge invariant object dressing the monopole with an operator that is odd with respect to the charge conjugation in $SO(N-2)$:
\be
(\mathfrak M^-)_{Q^{N-2}}\,\approx\, \mathfrak{M}^- \cdot \varepsilon_{i_i\dots i_{N-2}}Q^{i_1}\cdots Q^{i_{N-2}}\,
\ee
where the chiral fields are contracted using the Levi-Civita symbol of the residual $SO(N-2)$ factor, $\varepsilon$. This monopole operator is usually called \emph{baryon monopole}: let us observe that it only exists for $F\geq N-2$, it has non-trivial charge under both $\mathbb{Z}_2^\cC$ and $\mathbb{Z}_2^\cM$ and transforms in the rank-$(N-2)$ antisymmetric representation of $SU(F)$.

Another type of operator is relevant for us, having non-trivial magnetic fluxes with respect to two different Abelian factors in $U(1)^r$. Semi-classically, it can be written as:
\be
\mathfrak M^{\div}\,\approx\, \text{exp}\left(\frac{\alpha_1-\alpha_2}{g^2}+i(\phi_1-\phi_2)\right)\,.
\ee
where the two lined up bullets denote the fact that two different fluxes are turned on. Such monopole breaks the $SO(N)$ gauge group down to $S(O(N-4)\times O(4))$,\footnote{Actually, the $SO(4)$ factors is further broken to $U(2)$.} and it is not gauge invariant unless dressed with a conjugation-odd operator in $SO(N-4)$. This can be done defining the gauge invariant operator $\varepsilon\mathfrak M^\div_{Q^{N-4}}$,\footnote{The $N-4$ chiral fields dressing the monopoles are contracted with the Levi-Civita symbol of the $SO(N-4)$ residual group.} existing only for $F\geq N-4$. In theories with only one gauge group factor, $\mathfrak M^{\div}$ is not really a chiral operator; however, it plays a crucial role in dualities between orthogonal quiver theories discussed in this paper. 

In the theory under consideration, the last operator that deserve to be mentioned is the usual baryon:
\be
\label{eq:BaryonDef}
\cB\,=\, \varepsilon_{i_1\dots i_N}\,Q^{i_1}\cdots Q^{i_N}\,.
\ee
In the main text, different baryon-like operators can appear; in that case, we will denote them by an $\varepsilon$ followed by the fields contracted with the Levi-Civita symbol: for instance, the baryon in \eqref{eq:BaryonDef} could be also denoted by $\varepsilon Q^N$.

Once we have understood which kind of operators can be part of the chiral ring in 3d SQCD with orthogonal groups, we can easily discuss the Seiberg-like duality proposed by Aharony, Razamat, Seiberg and Willett (ARSW) in \cite{Aharony:2013kma}. The theory dual of $\CT_A$, $SO(N)$ SQCD with $F$ flavors $Q$, $\CW_{\CT_A}=0$, is $\CT_B$,  $SO(F-N+2)$ gauge theory with $F$ flavors $q$, $F(F+1)/2$ singlets $M_{ij}$ transforming in the symmetric representation of $SU(F)$ and superpotential:
\be
\cW_{\CT_B}\,=\, \sigma \mathfrak{M}^+\,+\, M_{ij}{\Tr} (q^iq^j)\,.
\ee
The map of the chiral ring generators is the following:
\be
\begin{tabular}{c c c }
$\mathcal T_A$ & & $\mathcal T_B$ \\
\hline
$\Tr(Q_i Q_j)$ & & $M_{ij}$  \\
$\mathfrak M^+$ & & $\sigma$  \\
$(\mathfrak M^-)_{Q^{N-2}}$ & & $\varepsilon q^{N-F+2}$  \\
$\varepsilon Q^{N}$ & & $(\mathfrak M^-)_{q^{N-F}}$  \\
\end{tabular}
\ee
Observe that baryons and baryon-monopoles are mapped to each other.

\subsection{$O(N)_\pm\,,\,\mathrm{Pin}(N)\text{ and }\mathrm{Spin}(N)$}
Different gaugings of charge conjugation and the magnetic $\mathbb{Z}_2^\cM$ discrete symmetry leads to different gauge groups, enjoying the same algebra as $SO(N)$ but differing in their global properties; in particular, the spectrum of chiral operators will be different.
\begin{itemize}
\item The gauge group $O(N)_+$ is obtained gauging $\mathbb{Z}_2^\cC$, {\it i.e.} the orthogonal reflection. Such $O(N)$ group is the most common in literature: the gauging of charge conjugation makes the baryon and the baryon-monopole not gauge invariant and they are not part of the chiral ring anymore.

\item If we gauge the diagonal combination $(\mathbb{Z}_2^\cC\times \mathbb{Z}_2^\cM)/\mathbb{Z}_2$, the less common $O(N)_-$ group is obtained; in this theory, only operators which are even (odd) under both charge conjugation and $\mathbb{Z}^\cM_2$ symmetry are gauge invariant: for this reason the monopole $\mathfrak{M}^+$ and the baryon $\cB$ are both projected out, while the baryon-monopole $(\mathfrak M^-)_{ Q^{N-2}}$ survives. However, the monopole usually denoted as $\mathfrak{M}^+_{\mathrm{Spin}}$, having twice the minimal flux, survives. 

\item $\mathrm{Spin}(N)$ theories are built gauging $\mathbb{Z}_2^\cM$. The (baryon-)monopole is projected out but the monopoles with double fluxes, $\mathfrak M^+_{\mathrm{Spin}}$ and $(\mathfrak M^-_{\mathrm{Spin}})_{Q^{N-2}}$, are still chiral operators on the Coulomb branch.

\item Finally, $\text{Pin}(N)$\footnote{To be precise, there exists two versions of $\text{Pin}(N)$: 
$\text{Pin}^\pm(N)$, as discussed in \cite{Cordova:2017vab}.} theories are obtained gauging both the discrete global symmetries; the Coulomb branch is parametrized by $\mathfrak{M}_{\mathrm{Spin}}^+$ while all baryonic-like operators (including monopoles) are projected out.
\end{itemize}

All such theories enjoy Seiberg-Like duality similar to the ARSW duality \cite{Aharony:2011ci,Aharony:2013kma}. $O(N)_+$ SQCD with $F$ flavors is dual to $O(F-N+2)_+$ SQCD with $F$ flavors, $N(N+1)/2$ $M_{ij}$ singlets duals of the meson ${\Tr} Q_i Q_j$, the singlet $\sigma$ dual of $\mathfrak M^+$ and the usual superpotential $\cW=\sigma \mathfrak M^++{\Tr}(q^i M_{ij}q^j)$; an analogous duality holds for $\text{Pin}(N)$ SQCD. Finally, $O(N)_-$ SQCD is dual to $\text{Spin}(F-N+2)$ SQCD (with singlets and appropriate superpotential): further details about the chiral ring mapping can be found in \cite{Aharony:2013kma}.
\bibliographystyle{ytphys}
\bibliography{ref}
\end{document}